\documentclass[aps,preprint,superscriptaddress,amsmath,amssymb,floatfix,longbibliography,titlepage,dcolumn]{revtex4}
\usepackage{dcolumn}   
\usepackage{bm}        
\usepackage{amssymb}   

    \usepackage{textcomp}
    \usepackage[utf8]{inputenc}
    \usepackage{graphicx}
    \usepackage{xcolor}
    \usepackage{siunitx}
    \usepackage{hyperref}
    \usepackage[normalem]{ulem}
    \usepackage{here}
    \usepackage{cancel}

\usepackage{bm}

\begin{document}

\title{Disentangling hierarchical relaxations in glass formers \\ via dynamic eigenmodes}
\author{Wensi Sun}
\affiliation{Department of Physics and State Key Laboratory of Surface Physics, Fudan University, Shanghai, 200438, P. R. China}
\author{Yanshuang Chen}
\affiliation{Department of Physics and State Key Laboratory of Surface Physics, Fudan University, Shanghai, 200438, P. R. China}
\author{Wencheng Ji}
\affiliation{Department of Physics of Complex Systems, Weizmann Institute of Science, Rehovot, 234 Hertzl St., Israel}
\author{Yi Zhou}
\affiliation{Department of Physics and State Key Laboratory of Surface Physics, Fudan University, Shanghai, 200438, P. R. China}
\author{Hua Tong}
\affiliation{Department of Physics, University of Science and Technology of China, Hefei, 230026, P. R. China}
\author{Ke Chen}
\affiliation{Institute of Physics, Chinese Academy of Sciences, Beijing, 100190, P. R. China}
\author{Xiaosong Chen}
\affiliation{School of Systems Science, Beijing Normal University, Beijing, 100878, P. R. China}
\author{Hajime Tanaka}
\email{tanaka@iis.u-tokyo.ac.jp}
\affiliation{Research center for Advanced Science and Technology, The University of Tokyo, 4-6-1 Komaba, Meguro-ku, Tokyo 153-8505, Japan}
\affiliation{Department of Fundamental Engineering, Institute of Industrial Science, The University of Tokyo, 4-6-1 Komaba, Meguro-ku, Tokyo 153-8505, Japan}
\author{Peng Tan}
\email{tanpeng@fudan.edu.cn}
\affiliation{Department of Physics and State Key Laboratory of Surface Physics, Fudan University, Shanghai, 200438, P. R. China}
\affiliation{Institute for Nanoelectronic Devices and Quantum Computing, Fudan University, Shanghai, 200438, P. R. China}


\begin{abstract}
{\bf Hierarchical dynamics in glass-forming systems span multiple timescales, from fast vibrations to slow structural rearrangements, appearing in both supercooled fluids and glassy states. Understanding how these diverse processes interact across timescales remains a central challenge. Here, by combining direct particle-level observations with a dynamic eigenmode approach that decomposes intermediate-timescale responses into distinct modes, we reveal the microscopic organisation of relaxation dynamics in two-dimensional colloidal systems. We identify five classes of modes characterizing hierarchical dynamics: (i) quasi-elastic modes, (ii) slow-reversible string modes contributing to dynamic heterogeneity, (iii) slow-irreversible string modes leading to flow, (iv) fast-$\beta$ modes with fast-reversible strings, and (v) random noise modes. The emergence of quasi-elastic modes marks the onset of glassy dynamics, while reversible string modes dominate dynamic heterogeneity throughout both supercooled and glassy regimes. Our findings offer a unified microscopic framework for understanding how distinct relaxation processes interconnect across timescales, illuminating the mechanisms driving glass formation.}
\end{abstract}

\date{\today}
\maketitle
%
%


\section*{Introduction}

The onset of hierarchical dynamics is one of the hallmarks of glass transition approaches, universally observed in various glassy systems ~\cite{chandler2010dynamics,keys2011excitations,yu2014beta,guiselin2022microscopic,scalliet2022thirty,debenedetti2001supercooled,stillinger1995topographic,bouchaud2024dynamics}. Experimental studies using dielectric, mechanical, and light-scattering spectroscopy have resolved the hierarchical dynamics as characteristic peaks or an excess wing in the relaxation spectra~\cite{korber2020systematic,schmidtke2013reorientational,hecksher2017toward}, which refer to different types of motions, including fast vibrations, fast-$\beta$, slow-$\beta$, and $\alpha$ relaxations~\cite{guiselin2022microscopic,scalliet2022thirty,yu2014beta}. Revealing the microscopic behaviours of these distinct motions, especially probing how they organise into hierarchical dynamics across multiple timescales, is crucial to understanding physical properties such as viscoelasticity, stability~\cite{yu2015suppression,rodriguez2022ultrastable}, and fragility~\cite{chen2023visualizing}.

A fundamental framework for understanding hierarchical dynamics is the potential energy landscape (PEL) picture~\cite{stillinger1995topographic,debenedetti2001supercooled}, which suggests that glassy systems have a rugged landscape composed of numerous metastable states. Near the glass transition, the dynamics involve two main components: fast vibrational motions within individual basins, captured by vibrational modes, and activated jumps between basins, which give rise to a range of intermediate relaxation behaviours bridging fast-$\beta$ and $\alpha$ processes. These inter-basin rearrangements typically involve particles localised in soft regions associated with low-frequency vibrational modes~\cite{widmer2008irreversible,zhang2021fast,hu2022origin,widmer2009localized,ji2019theory}, and often exhibit string-like geometries~\cite{donati1998stringlike,zhang2021fast,starr2013relationship,gokhale2014growing,hima2015direct,stevenson2006shapes,yip2020direct,pica2024local}. Although these strings share similar structural features, they represent distinct dynamic behaviours occurring on different timescales and often involve overlapping sets of particles~\cite{widmer2008irreversible}. Their nonlinear interactions and spatial entanglement lead to the emergence of dynamic heterogeneity, particularly after the accumulation of rearrangements over the $\alpha$-relaxation timescale~\cite{keys2011excitations,mishra2014dynamical,dauchot2005dynamical,chandler2010dynamics,berthier2011dynamical,ishino2025microscopic,tanaka2025structural}. Even deep in the glassy state, string-like patterns persist within vibrational modes around the boson peak~\cite{hu2022origin,jiang2024stringlet}, where an excess in the vibrational density of states $D(\omega)/\omega^{d-1}$ signals the breakdown of Debye scaling.

The diversity and complexity of microscopic behaviours in the intermediate-relaxation regime pose a major challenge to achieving a comprehensive understanding of hierarchical dynamics across timescales. This regime, which spans a broad range of durations, encompasses both nonlinear elastic responses and a variety of structural rearrangement events --- each of which can significantly influence the long-time glassy dynamics. Disentangling and identifying these distinct processes is particularly difficult in experimental systems, where direct access to the underlying mechanisms is limited.

Here, we demonstrate that the dynamic eigenmode approach (DEA) offers a powerful means to address this challenge. 
DEA analyses the short-time displacements $\Delta\mathbf{r}$ of particles sampled over a long observation window $t_\mathrm{a}$ that spans the $\alpha$-relaxation regime. From these displacements, it constructs a time-averaged correlation matrix $\mathbf{D}_{ij}(\Delta t, t_\mathrm{a}) = \langle \Delta\mathbf{r}_i \cdot \Delta\mathbf{r}_j \rangle_{t_\mathrm{a}}$, where $\Delta\mathbf{r}_i \cdot \Delta\mathbf{r}_j$ captures the correlation between the motions of particles $i$ and $j$ over a short interval $\Delta t$, incorporating both vibrational and jump-like components. Diagonalising $\mathbf{D}$ yields a set of dynamic modes: the eigenvectors represent spatial patterns of collective motion, while the eigenvalues quantify their contribution to the dynamics on the $\Delta t$ timescale. By separating temporally distinct behaviours into orthogonal components, DEA provides a mode-resolved view of the intermediate relaxation regime. Importantly, it is directly applicable to both simulations and experiments with single-particle resolution, such as colloidal systems~\cite{weeks2000three,zheng2011glass,chen2023visualizing}.

Although DEA is rooted in linear decomposition, the resulting modes may be interpreted as effective, time-averaged generalisations of vibrational eigenmodes, renormalised by thermal fluctuations and anharmonic effects. As such, they reflect the time-dependent spatiotemporal organisation of mobility within the disordered structure~\cite{henkes2012extracting}.

By applying DEA to two-dimensional (2D) colloidal experiments, and validating the results with numerical simulations, we uncover microscopic insights into the hierarchical dynamics near the glass transition. 
To maintain clarity and avoid overlap with conventional spectral terminology, we refer to the reversible and irreversible string-like excitations resolved by DEA as intermediate modes, rather than ``slow-$\beta$'' modes, emphasizing their role across the fast $\beta$-to-$\alpha$ relaxation crossover without implying specific spectral signatures.
We identify five distinct classes of modes that characterise the complex intermediate-timescale dynamics: (i) quasi-elastic modes, (ii) modes with slow-reversible strings that contribute to dynamic heterogeneity, (iii) modes with slow-irreversible strings that evolve into conventional flow, (iv) fast-$\beta$ modes composed of fast-reversible strings, and (v) random noise modes. The emergence of quasi-elastic modes [(i)] signals the onset of the glass transition, while reversible string modes [(ii)] increasingly dominate over irreversible ones [(iii)] as the system becomes more deeply supercooled. These results reveal how distinct dynamic behaviours are interrelated and organised across timescales, offering a unifying microscopic perspective on the complex relaxation processes underlying glassy dynamics.

\subsection*{Colloidal systems with hierarchical dynamics}

We perform colloidal experiments using a 2D binary mixture of NIPAm colloids with diameters $\sigma_\mathrm{L}:\sigma_\mathrm{S}\approx1.3~\mathrm{\mu m}:1~\mathrm{\mu m}$ at $22~^\circ\mathrm{C}$ and number ratio  $N_\mathrm{L}:N_\mathrm{S}\approx1$ (Extended Data Fig.~\ref{fig:NIPAm}). The NIPAm colloidal particles interact via soft repulsions, and their size is temperature-sensitive~\cite{chen2011measurement}. By tuning the area fraction $\phi$ of colloids, we obtain samples ranging from a supercooled liquid near the onset point ($\phi_{\rm on}\approx0.74$)~\cite{tanaka2025structural} to a denser state approaching the glass transition ($\phi_{\rm g}\approx0.83$)~\cite{chen2010low}. After equilibrating the sample for $3$ hours, we record the particle dynamics using bright-field microscopy at $60$ frames per second (fps) over a duration of $1,000$~s, capturing trajectories of approximately 3,500 particles within the field of view. 

For comparison, we also use another charged colloidal system composed of monodispersed poly(methyl methacrylate) (PMMA) particles ($\sigma\approx 2$~$\mu$m), grafted with polyhydroxystearic acid (PHSA) (Extended Data Fig.~\ref{fig:PMMA}). These PMMA colloids are negatively charged and interact with a hard-core Yukawa potential with a screening length of $\approx 1$~$\mu$m~\cite{tan2012understanding}. To avoid crystallisation, we fix the packing fraction at $\phi = 0.25$, corresponding to the onset of slow dynamics ($\phi_{\rm on}$). Particle trajectories in both systems are extracted using established particle tracking techniques~\cite{crocker1996methods}.

In the NIPAm colloidal system close to $\phi_{\rm g}$, we observe that the single-particle movement over a timescale of $\tau_\alpha$ consists of cage-confined rattling characterised by frequent, instantaneous jumps that lead to structural rearrangements (Fig.~\ref{fig:Fig1}a(i)). Notably, these jumps often occur collectively among neighbouring particles in the form of string-like excitations (Fig.~\ref{fig:Fig1}a(ii)). These strings can be further categorised based on their amplitude and reversibility.

Interestingly, the relaxation spectrum in this colloidal system exhibits a hierarchical structure, closely resembling those reported in experimental studies of metallic~\cite{yu2014beta} and molecular glasses~\cite{guiselin2022microscopic,blochowicz2003susceptibility}, as shown in 
Fig.~\ref{fig:Fig2}a.  We obtain the relaxation spectrum by calculating the imaginary part of the dynamic susceptibility $\chi''(\tau^{-1})$ from the distribution of relaxation times $\tau$~\cite{guiselin2022microscopic,scalliet2022thirty,blochowicz2003susceptibility} (se e Methods). The positions of the $\alpha$ peaks in these spectra coincide with $\tau_\alpha$,  which is independently determined from the decay of the intermediate scattering function $F_s(q,t)$ (Extended Data Fig. \ref{fig:NIPAm}c, Methods). As $\phi$ increases towards $\phi_{\rm g}\approx0.83$, the $\alpha$ peak becomes increasingly asymmetric. A power-law excess wing, $\chi''(\tau^{-1}) \sim \tau^{0.21}$, appears on its high-frequency side, followed by a shoulder associated with intermediate-mode dynamics~\cite{lunkenheimer2002dielectric,tanaka2004origin,guiselin2022microscopic,scalliet2021excess,scalliet2019nature}. The dynamics in this timescale regime are sub-diffusive, as evidenced by the plateau and subsequent rise in the mean square displacement (MSD) (Extended Data Fig.~\ref{fig:NIPAm}b). Within the intermediate region, we observe several types of string-like motions. Below, we apply the dynamic eigenmode approach (DEA) to systematically examine their contributions to the hierarchical dynamics. Here, we note that due to the overdamped nature of the vibrational motion in our colloidal system, we do not detect the high-frequency peaks typically associated with inertial vibrations in atomic or molecular systems.


\subsection*{Dynamic eigenmode approach}

The dynamic eigenmode approach (DEA) analyses the trajectories of $N$ particles recorded over a time interval $t_\mathrm{a}$, typically chosen to be comparable to the structural relaxation time $\tau_\alpha$. The trajectory should be high-resolution, containing $M \gg dN$ frames ($d$ is the dimension of the system). We begin by calculating the particles' displacements  $\Delta \mathbf{r}(t)=\{\Delta\mathbf{r}_1(t),\Delta\mathbf{r}_2(t),...,\Delta\mathbf{r}_N(t)\}^\mathrm{T}$ during an interval of $\Delta t$, where $\Delta \mathbf{r}_i(t) = \mathbf{r}_i(t + \Delta t) - \mathbf{r}_i(t)$, and $\mathbf{r}_i(t)$ denotes the position of particle $i$ at time $t$. By sweeping $t$ from the first to the final frame ($t \in [1, M]$), we obtain a time series $\Delta \mathbf{r} = \{\Delta\mathbf{r}(1), \Delta\mathbf{r}(2), ..., \Delta\mathbf{r}(M)\}$ that contains detailed information on when and where dynamic behaviours, such as collective jumps and vibrational motions, occur within $t_\mathrm{a}$.

Finally, we construct a $dN \times dN$ correlation matrix $\mathbf{D} = \frac{1}{M} \Delta \mathbf{r} \cdot \Delta \mathbf{r}^\mathrm{T}$, where each element is given by $\mathbf{D}_{ij} = \langle \Delta \mathbf{r}_i(t) \cdot \Delta \mathbf{r}_j(t) \rangle_{t \in [1, M]}$. In our study, to unravel the characteristic motions linked to hierarchical dynamics in colloidal systems, $\Delta t$ is carefully chosen within the intermediate-relaxation regime, and $t_\mathrm{a}$ is set to span several $\tau_\alpha$. With this configuration, the matrix $\mathbf{D}$ captures both structure rearrangement events, such as string-like excitations, and cooperative in-cage motions that persist over long timescales, such as slow and large-scale elastic fluctuations.

We then perform eigen-decomposition of the matrix $\mathbf{D}$, obtaining $dN$ dynamic modes that characterise the intermediate-timescale responses (intermediate modes) in terms of their eigenvalues $\lambda$ and eigenvectors $\mathbf{e}_\lambda$. The eigenvalue $\lambda$, which has the same units as $a^2$ (with $a$ being the average inter-particle distance), quantifies the contribution (or weight) of each eigenmode to the dynamics at the timescale $\Delta t$, satisfying the relation $\Sigma\lambda = \langle(\Delta \mathbf{r}(\Delta t))^2\rangle$ (see Methods for details). Modes with larger $\lambda$ capture more significant dynamic responses over $\Delta t$, while the corresponding eigenvectors $\mathbf{e}_\lambda$ illustrate the spatial organisation of collective particle motions.

Thus, the spectrum of $\lambda$ and the structure of the vector fields $\mathbf{e}_\lambda$ together represent the diverse dynamic behaviours emerging on the $\Delta t$ timescale. For more effective characterisation, we introduce a dimensionless parameter $\nu = a^2 / \lambda$ to relabel the eigenmodes and compute the accumulative number of states (ANOS) spectrum, $N(\nu)$ (Methods). Additionally, we project the long-time trajectories associated with $\alpha$-relaxation onto these intermediate modes, enabling us to explore the connection between long-timescale relaxation and various intermediate-timescale dynamical processes. This projection allows us to probe how relaxation dynamics evolve and self-organise hierarchically as the system approaches the glass transition.

From a theoretical perspective, the dynamic eigenmodes obtained from $\mathbf{D}$ can be regarded as finite-temperature, anharmonic generalisations of the harmonic vibrational modes derived from the system's Hamiltonian. The thermal fluctuations and anharmonic interactions act as perturbations to the ideal harmonic system, leading to mode mixing, renormalisation, and temporal evolution. Here, we perform the numerical simulations, and compare the eigenmodes of $\mathbf{D}$ with those of the covariance matrix $\mathbf{C}_{ij} = \langle [\mathbf{r}_i(t) - \bar{\mathbf{r}}_i] \cdot [\mathbf{r}_j(t) - \bar{\mathbf{r}}_j] \rangle$ (where $\bar{\mathbf{r}}_i$ denotes the equilibrium position of particle $i$) and the vibrational modes(VM) from Hessian matrix $\mathbf{H}$~(Methods; Extended Data Fig.~\ref{fig:simualtion} and \ref{fig:perturbation}).

For low-temperature glasses in the harmonic limit ($T \to 0$), eigenmodes of $\mathbf{D}$ under long-time averaging are static and converge to those of $\mathbf{C}$. The eigenmode can be seen as the combination of several VM, with its frequency slightly shifting from that of VM (Extended Data Fig.~\ref{fig:perturbation}a and c). However, when approaching $T_\mathrm{g}$, the growing nonlinear effect induces the cooperative activation of vibrational modes with different frequencies, resulting in extended collective motions and structural rearrangements with characteristic timescales longer than the vibrational ones (Extended Data Fig.~\ref{fig:perturbation}b). As the system explores different metastable states, the eigenmodes of $\mathbf{C}$ may reflect collective relaxations or flow-like behaviours with diverging eigenvalues. Meanwhile, the eigenmodes of $\mathbf{D}$ represent effective, time-dependent collective modes, encoding both localised rearrangement events and persistent in-cage fluctuations (Extended Data Fig.~\ref{fig:simualtion}). This interpretation provides a physical basis for the observed temporal variability of the modes and supports the use of DEA as a framework for capturing thermally activated, multi-timescale nonlinear dynamics beyond the harmonic regime.


\subsection*{Classification of dynamic eigenmodes}

\begin{figure*}[htpb]
\includegraphics[width=0.75\linewidth]{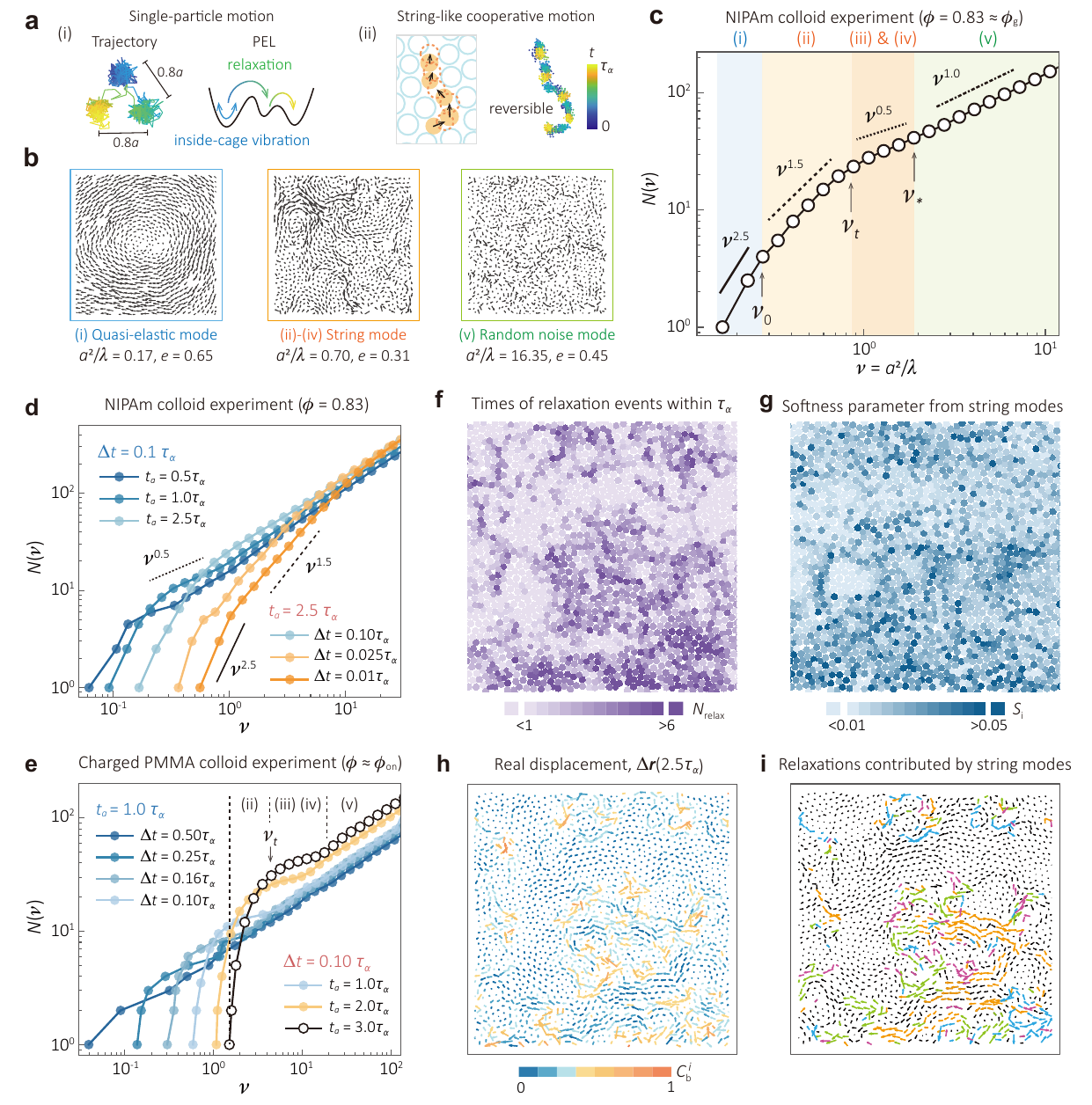}
\caption{\textbf{Dynamic eigenmodes of intermediate relaxations in colloidal systems.} 
$\mathbf{a}$, Typical particle trajectories in our colloidal system at $\phi_\mathrm{g}\approx0.83$. Inside-cage rattlings mix with instantaneous jumps in a string-like manner. 
$\mathbf{b}$, Polarisation vector $\mathbf{e}_\nu$ of distinct types of dynamic eigenmodes at $\phi_\mathrm{g}$ in NIPAm colloidal system, obtained by diagonalizing the matrix $\mathbf{D}$, $\Delta t = 0.1 \tau_\alpha$ and $t_{\rm a} = 2.5 \tau_\alpha$.
$\mathbf{c}$, Accumulated number of states (ANOS) spectrum of the modes in $\mathbf{b}$. $\nu_0$, $\nu_t$ and $\nu_*$ are the turning points of distinct $\nu^k$ regions. 
$\mathbf{d}$-$\mathbf{e}$, ($\Delta t$, $t_{\rm a}$) dependence of ANOS spectra in NIPAm system near $\phi_{\rm g}$ ({\bf d}, $\phi = 0.83$) and charged PMMA system near $\phi_\mathrm{on}$ ({\bf e}). $k$ varies with $\phi$ and interaction potentials, but remains independent of $\Delta t$ and $t_\mathrm{a}$. Note that there is no quasi-elastic modes (i) in PMMA system. $\mathbf{f}$, Average number of relaxation events within $\tau_\alpha$, $N_\mathrm{relax}$. A single relaxation event for particle $i$ is identified if at least one of its neighbours changes.
$\mathbf{g}$, Softness parameter $S_i$ calculated from string modes (ii)-(iv).
$\mathbf{h}$, Displacement $\Delta \mathbf{r}$ during 2.5 $\tau_\alpha$. Vectors are coloured based on the ratio of neighbouring-particle change, $C_b^i$.
$\mathbf{i}$, Displacement in $\mathbf{i}$ projected to string modes (ii)-(iv). The coloured vectors correspond to strings captured in different string modes shown in Extended Data Fig.~\ref{fig:NIPAm}d. 
\label{fig:Fig1}
}
\end{figure*}

Through DEA, we characterise the complex intermediate-timescale dynamic motions in our NIPAm colloidal system close to the glass transition ($\phi = 0.83$, and $t_{a} = 2.5\tau_\alpha$, $\Delta t = 0.1\tau_\alpha$). By combining the spatial features of $\mathbf{e}_\nu$ shown in Fig.~\ref{fig:Fig1}b with the ANOS spectrum in Fig.~\ref{fig:Fig1}c, we distinguish five types of modes: (i) quasi-elastic modes with an extended background ($\nu < \nu_0$), (ii) modes with slow-reversible strings ($\nu_0 < \nu < \nu_\mathrm{t}$), (iii) modes with slow-irreversible strings ($\nu \approx \nu_\mathrm{t}^+$), (iv) modes with highly-reversible fast strings ($\nu_\mathrm{t} < \nu < \nu_\ast$), and (v) random-like noise modes ($\nu > \nu_\ast$). 

Each mode type occupies a characteristic region on the ANOS spectrum and exhibits distinct power-law slopes and crossover points in the log-log representation (Fig.~\ref{fig:Fig1}c). We estimate these slopes using $N(\nu) \sim \nu^k$ and define the crossover points as $\nu_0$, $\nu_t$, and $\nu_\ast$, corresponding to transitions between regions of different scaling exponents. For comparison, phonon modes in crystalline solids follow the Debye scaling $N(\nu) \sim \nu^{d/2}$ on the ANOS spectrum. Figure~\ref{fig:Fig1}d shows that for $\Delta t$ in the intermediate relaxation timescale and $t_a$ around $\tau_\alpha$, the exponent $k$ can be reliably identified across various choices of matrix parameters. The number of extracted modes within each class depends on these parameters (Extended Data Fig.~\ref{fig:Si}a).

The quasi-elastic modes in (i) exhibit an extended, vortex-like structure (Fig.~\ref{fig:Fig1}b, left) and follow a scaling exponent of $k = 2.5$ on the ANOS spectrum (blue region in Fig.~\ref{fig:Fig1}c). These modes represent quasi-elastic responses of the system near and below $\phi_\mathrm{g}$ (or $T_\mathrm{g}$) on intermediate-relaxation timescales~(much larger than the oscillation timescale, as shown in Extended Data Fig.~\ref{fig:perturbation}). As evidence, we find that such modes are absent in supercooled NIPAm samples when $\phi \leq 0.80$ (Fig.~\ref{fig:Fig2}b), as well as in the monodisperse PMMA system near $\phi_\mathrm{on}$ (Fig.~\ref{fig:Fig1}e). Furthermore, they disappear when the dynamic eigenmode analysis is performed using a large interval $\Delta t \approx \tau_\alpha$.

Importantly, whenever these extended modes are present, we consistently observe the $k = 2.5$ scaling in 2D glassy systems, independent of interaction potential --- for instance, in simulations of the WCA binary system (Extended Data Fig.~\ref{fig:simualtion}e-f). Previous studies have predicted a similar $k = 2.5$ scaling for the ANOS vibrational spectrum $N(\omega^2)$ associated with quasi-localised vibrational modes in models incorporating anharmonic effects~\cite{ji2019theory,ji2020thermal}. In our system, the modes in (i) likely correspond to quasi-elastic responses that emerge near or below $\phi_\mathrm{g}$, where quasi-localised vibrational modes hybridise with extended plane-wave-like excitations under the influence of anharmonicity. 
The emergence of these modes signals a mechanically driven self-organisation that sets in as the system surpasses $\phi_\mathrm{g}$~\cite{yanagishima2017common,Tong2020}.
These modes do not trigger structural rearrangements but contribute significantly to the mean squared displacement (as reflected by their large $\lambda$), suggesting a high energy barrier for $\alpha$ relaxation along these directions.

For the modes in (ii), (iii), and (iv), the corresponding vector fields exhibit extensive string-like features involving a small number of particles (Fig.~\ref{fig:Fig1}b, middle). These modes fall within an intermediate range of $\nu$ and can be broadly divided into two regions separated by $\nu_t$, each characterised by a different scaling exponent $k$ on the ANOS spectrum: approximately $k = 1.5$ for $\nu_0 \leq \nu \leq \nu_t$ and $k = 0.5$ for $\nu_t \leq \nu \leq \nu_\ast$ (Fig.~\ref{fig:Fig1}c). It is important to note that $k$ is not a universal quantity --- it varies across systems depending on particle composition, interaction potential, and packing fraction, reflecting the system-specific nature of relaxation dynamics (Fig.~\ref{fig:Fig1}d-e and Fig.~\ref{fig:Fig2}b). 
Modes near $\nu_t$ are only weakly quasi-localised, with a participation ratio around $e \approx 0.3$, and give rise to an excess peak in the reduced density of states (DOS) spectrum, as illustrated in Extended Data Fig.~\ref{fig:basic}.

Surprisingly, although string modes constitute only a small fraction (approximately 1\%) of all $dN$ eigenmodes, they effectively capture the relaxation behaviours and play a dominant role in structural relaxations on long timescales around $\tau_\alpha$. The average number of relaxation events per particle, $N_{\rm relax}$, within $\tau_\alpha$ --- defined by at least one neighbour change and shown in Fig.~\ref{fig:Fig1}f --- is well captured by the softness parameter $S_i$ derived from the string modes (Fig.~\ref{fig:Fig1}g), where $S_i = \sum_{\nu_0 \leq \nu \leq \nu_\ast} \lvert \mathbf{e}_\nu^i \rvert^2$, and $\mathbf{e}_\nu^i$ is the polarisation vector of particle $i$ in eigenmode $\mathbf{e}_\nu$. 
Moreover, the string modes account for nearly all significant structural rearrangements observed over a period of $2.5\tau_\alpha$. This is validated by comparing the actual displacement field $\Delta\mathbf{r}(2.5\tau_\alpha)$ (Fig.~\ref{fig:Fig1}h) with the weighted accumulative vector field reconstructed from the string modes, defined as $\Delta\mathbf{r}_\mathrm{modes} = \sum_{\nu_0 \leq \nu \leq \nu_\ast} (\mathbf{e}_\nu \cdot \Delta\mathbf{r}) \mathbf{e}_\nu$ (Fig.~\ref{fig:Fig1}i). The high degree of similarity between the two fields demonstrates the strong predictive power of string modes.
Since modes in (ii), (iii), and (iv) all feature string-like excitations as elementary units of structural relaxation, we further examine their respective roles in facilitating these relaxation processes.

We assess the reversibility of the modes using a mode-level irreversibility parameter $R_\nu$. For a given mode $\nu$, we project the short-time displacement vector $\Delta\mathbf{r}$ onto the corresponding eigenvector $\mathbf{e}_\nu$, defining the projection coefficient as $c_\nu(t, \delta t) = \mathbf{e}_\nu \cdot \Delta\mathbf{r} / \lvert \Delta \mathbf{r} \rvert$, where $\Delta \mathbf{r} = \mathbf{r}(t + \delta t) - \mathbf{r}(t)$ and $\delta t = \Delta t$. Since the eigenmodes are orthonormal, the set of coefficients ${c_\nu}$ satisfies the normalisation condition $\sum_\nu c_\nu^2 = 1$, with $c_\nu$ representing the contribution of mode $\nu$ to the displacement $\Delta\mathbf{r}$. A positive $c_\nu(t, \delta t)$ indicates motion along the direction of $\mathbf{e}_\nu$, while a negative value implies motion in the opposite direction.
To quantify the time-asymmetry of these projections, we define the irreversibility parameter as $R_\nu = \lvert \frac{\sum_t c_\nu(t, \delta t) \cdot \lvert c_\nu(t, \delta t) \rvert}{\sum_t c_\nu^2(t, \delta t)} \rvert$, which ranges from $0$ to $1$. A larger $R_\nu$ signifies greater irreversibility of the mode, with $R_\nu = 1$ corresponding to completely irreversible motion, and $R_\nu = 0$ indicating fully reversible dynamics. Illustrative examples of this metric are provided in Extended Data Fig.~\ref{fig:reversibility}a-b.

As shown in Fig.~\ref{fig:Fig2}c, the modes with $\nu < \nu_t$ in region (ii) exhibit a high degree of reversibility, with $R_\nu$ typically below 0.3. Just above $\nu_\mathrm{t}$, the modes in (iii) become highly irreversible, with $R_\nu > 0.65$. Further along the spectrum, modes near $\nu_\ast$ in region (iv) again become highly reversible, approaching $R_\nu \to 0$. Owing to these distinct reversibility characteristics, the three mode types play different roles in the manifestation of microscopic relaxation events.
We find that the string-like excitations in (ii) are associated with large-amplitude, instantaneous hopping motions, which can be reversible (Fig.~\ref{fig:Fig2}d and Extended Data Fig.~\ref{fig:reversibility}c). However, the reversibility is not absolute, particularly when these strings interact with neighbouring ones.

In contrast, the irreversible modes in (iii) exhibit flow-like behaviour. A comparison between the direction of $\mathbf{e}_\nu$ and the actual displacement fields $\Delta\mathbf{r}$ reveals that particles involved in strings from modes in (iii) move nearly unidirectionally over long timescales of several $\tau_\alpha$, even though their amplitudes over the short interval $\Delta t$ remain small (Fig.~\ref{fig:Fig2}e; see also Extended Data Fig.~\ref{fig:reversibility}d). These flow-like motions are sensitive to $\phi$, showing increased irreversibility (larger $R_\nu$) and smaller $\nu$ as $\phi$ decreases toward $\phi_\mathrm{on}$ (Fig.~\ref{fig:Fig2}c).
In regime (iv), the highly reversible modes near $\nu_\ast$ exhibit very small spatial amplitudes but still retain string-like features. This class of motion corresponds to the well-known fast-$\beta$ relaxations~\cite{zhang2021fast,yu2014beta}.

Moreover, our analysis reveals that the presence of different dynamic modes depends sensitively on the degree of supercooling. In particular, quasi-elastic modes [(i)] emerge only near and below the glass transition, as evidenced by their absence in supercooled samples with $\phi \leq 0.80$. In contrast, slow-reversible string modes [(ii)], slow-irreversible string modes [(iii)], and fast-$\beta$ modes [(iv)] exist across both supercooled liquids and glassy states, though their relative contributions evolve as the system slows down. Random noise modes [(v)] are present throughout all regimes. This distinction highlights that the onset of quasi-elastic modes provides a microscopic signature of the glass transition, consistent with the emergence of shear elasticity arising from mechanical self-organisation~\cite{yanagishima2017common,Tong2020}. In contrast,  string-like excitations dominate the hierarchical dynamics over a broad range of timescales, spanning from the fluid to the deeply glassy phases.


\subsection*{Microscopic insights of hierarchy and dynamic heterogeneity}

Finally, we investigate how relaxation dynamics evolve and organise into a hierarchical structure by projecting long-time $\alpha$-relaxation trajectories onto the dynamic eigenmodes associated with intermediate relaxations. To do so, we vary the timescale $\delta t$ from $0.1$ to $2.0\tau_\alpha$, and compute the projection coefficient $c_\nu(t, \delta t)$ (Extended Data Fig.~\ref{fig:landscape}). The contribution of eigenmode $\nu$ to the dynamics at a given timescale $\delta t$ is then quantified by the parameter $C_\nu(\delta t) = \max ( \lvert c_\nu(t, \delta t) \rvert )$, which captures the maximum mode-level activation within that interval.

Figure~\ref{fig:Fig3}a shows the evolution of $C_\nu(\delta t)$ as a function of $\delta t$ in four NIPAm colloidal systems, spanning volume fractions from $\phi_g \approx 0.83$ to $\phi_{on} \approx 0.74$. As $\delta t$ increases, certain intermediate modes --- particularly those in regions (ii) and (iii) near $\nu_\mathrm{t}$ --- begin to dominate the long-time relaxation dynamics. This is evidenced by the emergence of several $c_\nu$ peaks below $\nu_\ast$. Meanwhile, quasi-elastic modes [(i)] contribute negligibly to the long-time relaxation dynamics in the supercooled regime, consistent with their emergence only near and below the glass transition.

To quantify the respective contributions of modes (ii) and (iii), we compare their peak heights, $C_\mathrm{peak}(\delta t)$, within the intervals $\nu_0 < \nu \leq \nu_\mathrm{t}$ and $\nu > \nu_\mathrm{t}$, respectively. The evolution of these peak values with increasing $\delta t$ is shown in Fig.~\ref{fig:Fig3}b. Here, $C_\mathrm{peak}(\delta t)$ is defined as $C_{\nu}(\delta t)_\mathrm{max} - \bar{C}_\mathrm{random}$, where $C_{\nu}(\delta t)_\mathrm{max}$ is the maximum value of $C_\nu(\delta t)$ within the corresponding range, and $\bar{C}_\mathrm{random}$ is the average over the random noise modes (v) with $\nu > \nu_\ast$.

Interestingly, $C_\mathrm{peak}(\delta t)$ in regime (ii) evolves non-monotonically: it initially increases, peaks around $0.75$ to $1.0\tau_\alpha$, and then declines (Fig.~\ref{fig:Fig3}a-b). At the microscopic level, reversible hopping strings occur individually on the intermediate timescale. As $\delta t$ increases, the activations of these hopping strings accumulate, and some evolve into irreversible structural rearrangements, leading to anomalous diffusion around the $\alpha$-relaxation timescale.
Consequently, the kinetics associated with modes in (ii) underlie the emergence of dynamic heterogeneity, as indicated by the similar non-monotonic trend in the four-point susceptibility $\chi_4(\delta t)$ (Fig.~\ref{fig:Fig3}d; Methods). In real space, Fig.~\ref{fig:Fig3}c presents the spatial distribution of the softness parameter $S_i = \sum_{\nu \leq \nu_\mathrm{t}} \lvert \mathbf{e}_\nu^i \rvert^2$ over three consecutive time intervals and the total. Here, $S_i$ is calculated using $\Delta t = 0.05\tau_\alpha$, and the ensemble averages are taken over $t_\mathrm{a} = [0, 0.5\tau_{\alpha}]$, $[0.5\tau_{\alpha}, \tau_{\alpha}]$, $[\tau_{\alpha}, 1.5\tau_{\alpha}]$, and $[0, 1.5\tau_{\alpha}]$, respectively. The locations of mobile regions (orange patches) suggest that subsequent structural rearrangements tend to occur near existing ones, accumulating into larger mobile clusters over time.

This observation aligns with the dynamic facilitation (DF) picture~\cite{chandler2010dynamics,keys2011excitations,berthier2011dynamical,gokhale2014growing}. We propose that the reversibility of modes in category (ii) diminishes as they become integrated into DF chains. In this context, newly triggered excitations locally reorganize the surrounding structure, altering the energy landscape and effectively lowering the barriers for subsequent relaxation events.

\begin{figure*}[htpb]
\includegraphics[width=\linewidth]{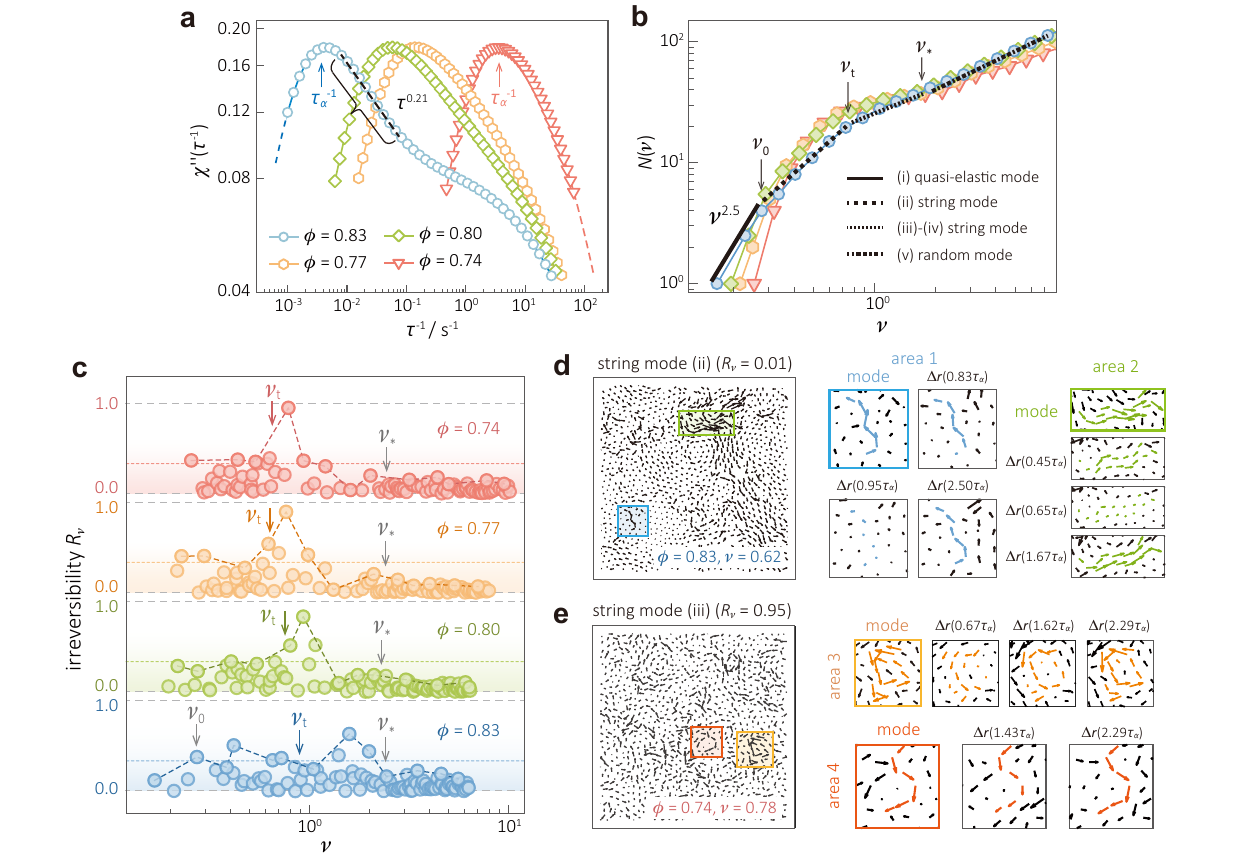}
\caption{\textbf{Characters of distinct intermediate modes towards the glass transition.} 
$\mathbf{a}$, Relaxation spectra characterised by the imaginary part of the dynamic susceptibility $\chi''(\tau^{-1})$. The blue curve ($\phi\approx\phi_\mathrm{g}$) indicates the typical hierarchical dynamics with a $\tau^{0.21}$ power-law excess wing followed by an intermediate-mode shoulder. 
$\mathbf{b}$, ANOS spectra towards the glass transition ($\Delta t=0.1\tau_\alpha$ and $t_\mathrm{a}=2.5\tau_\alpha$). The quasi-elastic mode associated with the $\nu^{2.5}$ region (i) is absent in systems with $\phi\leq0.80$. Note that $k$ is not well defined, as the $\nu_t$-$\nu$ slop varies with $\phi$ when $\phi<\phi_\mathrm{g}$.
$\mathbf{c}$, Irreversibility of modes characterised by parameter $R_\nu$. Larger $R_\nu$ indicates higher irreversibility. The coloured dashed lines refer to $R_\nu=0.3$.
$\mathbf{d}$-$\mathbf{e}$, Comparisons of $\mathbf{e}_\lambda$ and the displacement $\Delta \mathbf{r}$ at various t. Strings in a typical mode belonging to modes (ii) (in {\bf d}, $\phi=0.83\approx\phi_{\rm g}$) feature reversible string-like structure rearrangements. In contrast, strings in a typical mode belonging to modes (iii) (in {\bf e}, $\phi=0.74\approx\phi_\mathrm{on}$) feature irreversible unidirectional flow-like motions over long timescale of several $\tau_\alpha$.
\label{fig:Fig2}
}
\end{figure*}

\begin{figure*}[htpb]
\includegraphics[width=\linewidth]{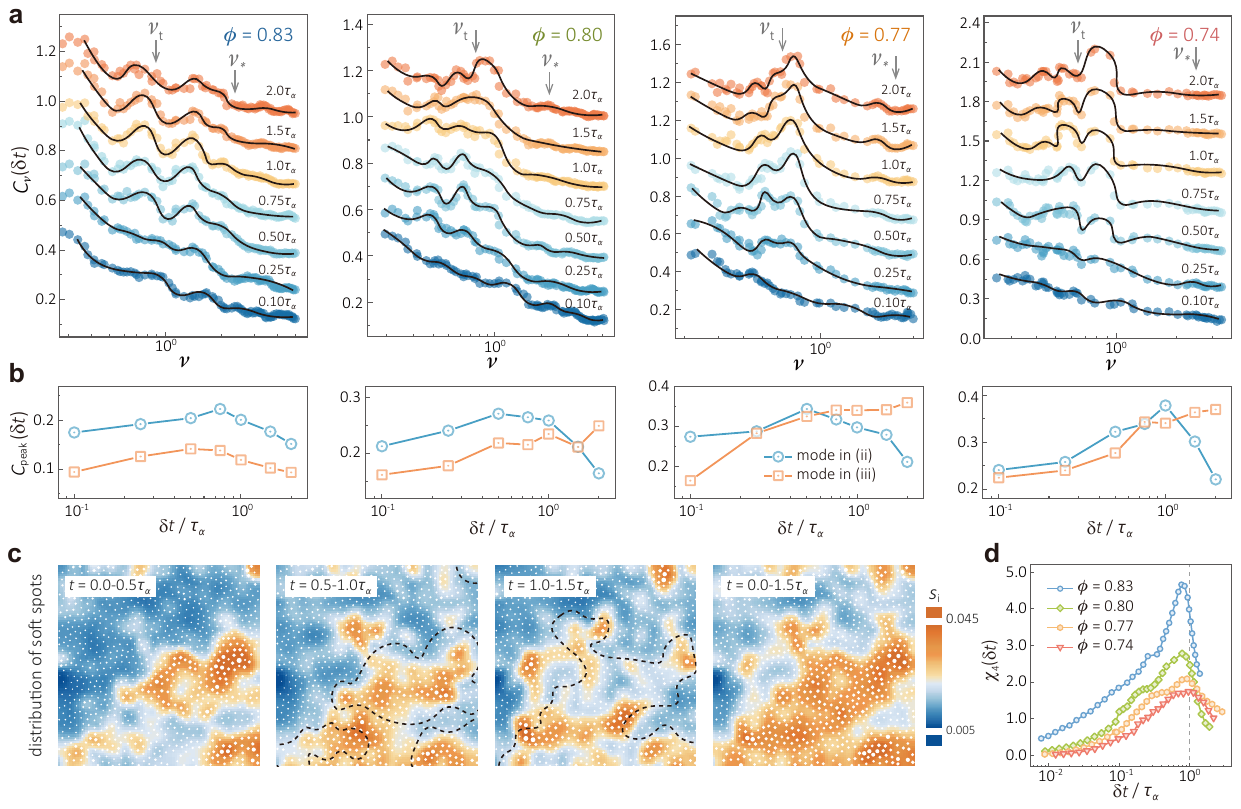}
\caption{\textbf{Contributions of distinct string modes to hierarchical and heterogeneous relaxations.} 
$\mathbf{a}$, Evolution of the contribution parameter $C_\nu(\delta t)$ with $\delta t$ towards glass transition in four NIPAm systems, from $\phi=0.83$ (left) to $0.74$ (right). Solid curves serve as the guidelines.
$\mathbf{b}$, Height of $C_\nu$ peaks associated with string modes (ii) and (iii) with respect to $\delta t$. $C_\mathrm{peak}(\delta t)$ of modes (ii) grows non-monotonically, peaking around $0.75$ to $1.0\tau_\alpha$. When $\phi\leq0.80$ where the flow-like motions dominate, $C_\mathrm{peak}(\delta t)$ of modes (iii) continuously increases with $\delta t$.
$\mathbf{c}$, Spatial distribution of $S_i$ from modes (ii) corresponding to three consecutive time periods ($[0,0.5\tau_\alpha], [0.5\tau_\alpha,1.0\tau_\alpha],[1.0\tau_\alpha,1.5\tau_\alpha]$) and the total $[0,1.5\tau_\alpha]$) at $\phi=0.83$. The colour of background and sizes of dots indicate $S_i$ calculated from modes (ii) ($\Delta t=0.05\tau_\alpha$). The new orange patches develop around the previous ones, indicating a dynamic facilitation phenomenon. 
$\mathbf{d}$, Dynamic heterogeneity characterised by the four-point susceptibility $\chi_4(\delta t)$. $\chi_4(\delta t)$ peaks around $\tau_\alpha$, similar to the $C_\mathrm{peak}(\delta t)$ in {\bf b}.
\label{fig:Fig3}
}
\end{figure*}

In contrast, the kinetics contributed by modes in (iii) evolve into conventional flow-like behaviour, similar to that observed in normal liquids. The $C_\nu$ peaks associated with these modes exhibit a nearly monotonic increase once they begin to dominate the relaxation dynamics at $\phi \leq 0.80$ (Fig.~\ref{fig:Fig3}b). On the relaxation spectrum, these modes contribute directly to the $\alpha$ peak, with only a subtle signature of intermediate-mode relaxations (Fig.~\ref{fig:Fig2}a).
For the modes in (iv), the values of $C_\nu(\delta t)$ remain small across all timescales, indicating that they do not play a significant role in the long-time structural relaxation. Accordingly, these modes are not directly relevant to the evolution of the long-time particle trajectories.


\section*{Discussion}

Using the dynamic eigenmode approach (DEA), we reveal how diverse intermediate-timescale dynamics serve as a bridge between fast-$\beta$ processes and long-time $\alpha$ relaxations in glassy systems. Although DEA is fundamentally a linear decomposition technique, it effectively captures the spatiotemporal signatures of emergent nonlinear excitations---including string-like rearrangements, quasi-elastic responses, and flow-like motions. This suggests that complex relaxation behaviours are mediated by low-dimensional, coherent structures embedded within the fluctuation field---structures that can be meaningfully accessed through linear projection. From a statistical physics perspective, the dynamic eigenmodes revealed by DEA can be interpreted as emergent, time-dependent collective excitations shaped by thermal fluctuations and anharmonic effects. In analogy with perturbation theory, they may be viewed as modified eigenmodes of the system’s Hamiltonian, renormalised by thermal noise and structural rearrangements. This interpretation supports the observed temporal variability of the modes as the system explores different basins of the potential energy landscape, highlighting DEA's value not only as a decomposition tool but also as a physically grounded lens onto finite-temperature dynamics in disordered systems.

Previous studies have attributed structural rearrangements in glasses to excitations of soft spots — localised regions linked to quasi-localised vibrational modes~\cite{li2022local,rainone2020pinching,manning2011vibrational,lerner2021low}. This framework offers a mechanical perspective, proposing that such rearrangements are governed by the self-organisation of mechanically stable networks embedded within the disordered matrix~\cite{yanagishima2017common,Tong2020}. Our results help refine this picture for finite-temperature systems, where thermal fluctuations and anharmonic effects play a critical role. Near $T_\mathrm{g}$ or $\phi_\mathrm{g}$, the localised soft spots originating from distinct vibrational modes and inherent PEL states may become collectively activated, giving rise to string-like motions through dynamic facilitation~\cite{chandler2010dynamics,keys2011excitations}, entropic effects~\cite{starr2013relationship,stevenson2006shapes}, or mechanical instability~\cite{yanagishima2017common}.
These intermediate relaxations may exhibit distinct reversibility and degree of collectivity due to the intrinsic differences among soft spots, and can be distinguished into modes (ii) and (iii) through DEA. Meanwhile, the existence of quasi-elastic modes (i) suggests that some low-frequency vibrational modes do not contribute to structural rearrangements due to high energy barriers, but instead sustain cooperative quasi-elastic responses at long timescales. They are reminiscent of the metastable dynamics with nonlinearity, and contribute to the mean square displacement in a power-law distribution as $N(\nu)\sim\nu^{2.5}$.


Note that the DEA results shown here are analyzed with $t_a$ around $\tau_\alpha$. Within this timescale, the configuration of glassy systems still preserves certain similarity with the initial state, ensuring the clear observation of independent collective motions in modes (ii) and (iii). When detecting with infinite $t_a$, we suggest that the quasi-elastic responses should persist with same power-law feature, while the signals of dynamic heterogeneity are averaged out. Although the vector fields of eigenmodes are not directly related to certain relaxation behaviours, the eigenvalues and spectrum will converge, reflecting the statistical features of nonlinear glassy dynamics that varies with systems and temperatures.

At the same time, DEA has inherent limitations. As a time-averaged, linear method, it may underrepresent strongly nonlinear or intermittent relaxation events that are crucial near the glass transition. Its sensitivity to analysis parameters (e.g., displacement interval $\Delta t$ and averaging window $t_\mathrm{a}$) requires careful tuning, and its empirical nature disconnects it from the system's Hamiltonian and intrinsic energy landscape. Moreover, mode interpretation can be nontrivial due to mixing between reversible and irreversible dynamics. These limitations point towards promising directions for future work, such as incorporating time-local correlation matrices or nonlinear dimensionality reduction techniques, to access even richer features of glassy relaxation. Nonetheless, DEA provides a robust and interpretable framework for dissecting hierarchical dynamics across timescales, with broad applicability to complex and disordered systems. We propose that eigenmode analysis --- by resolving spatial-temporal correlations in dynamic behaviour --- offers a powerful and broadly applicable framework for studying exotic dynamical phenomena in systems ranging from active matter to complex networks.



\clearpage

\newpage
\section*{Methods}
\subsection*{Colloidal experiments.}

Our experiments involve two types of colloidal systems with distinct interaction potentials. The first system consists of aqueous-based poly($N$-isopropyl acrylamide) (NIPAm) microgel colloidal spheres (Extended Data Fig.~\ref{fig:NIPAm}a(i)). The NIPAm particles interact via Hertzian short-range repulsion~\cite{chen2011measurement}. To prepare the 2D samples, a binary mixture of micron-sized NIPAm particles is confined between two coverslips to form a monolayer. To prevent crystallisation, the particle size and number ratios are fixed at $\sigma_\mathrm{L}:\sigma_\mathrm{S} = 1.3\ \mathrm{\mu m} : 1\ \mathrm{\mu m}$ (at $22~^\circ \mathrm{C}$) and $N_\mathrm{L} : N_\mathrm{S} = 1 : 1$, respectively. Samples are sealed with waterproof optical glue (Norland 63) to minimise evaporation.

We tune the packing fraction $\phi$ from $\phi = 0.74$ (supercooled fluid) to $\phi = 0.83$ (deeply supercooled, close to $\phi_{\rm g}$)~\cite{chen2010low} (Extended Data Fig.~\ref{fig:NIPAm}b-c). After equilibrating the sample for 3 hours, we image the dynamics using bright-field microscopy at 60 frames per second (fps) for 1,000 seconds, capturing trajectories of approximately 3,500 particles within the field of view. Trajectories are analysed using a standard particle-tracking algorithm~\cite{crocker1996methods}.

The second system is a non-equilibrium colloidal fluid composed of monodisperse poly(methyl methacrylate) (PMMA) particles grafted with polyhydroxystearic acid (PHSA) (Extended Data Fig.~\ref{fig:PMMA}). The particle diameter is $\sigma = 2\ \mathrm{\mu m}$ with polydispersity below 3\%. These particles are suspended in a solvent mixture of 1-iododecane and 1-iodododecane (volume ratio $\sim 1:4$), which provides both density and refractive index matching. The PMMA particles are negatively charged and interact via a hard-core Yukawa potential with a screening length of approximately $1\ \mathrm{\mu m}$~\cite{tan2012understanding}.

To avoid crystallisation, we fix the packing fraction at $\phi = 0.25$ (near $\phi_\mathrm{on}$). Particle motions are recorded via confocal microscopy at 20 fps over 400 seconds (spatial resolution $\sim$15 nm), and analysed using the same particle tracking method as for the NIPAm system.

\subsection*{Simulation.}

Our research also incorporates molecular dynamics (MD) simulations performed using the open-source software Large-scale Atomic/Molecular Massively Parallel Simulator (LAMMPS)~\cite{plimpton1995fast}. We adopt the well-established Kob-Andersen 2D binary mixture model, consisting of $N = 2{,}000$ classical particles with a number ratio of $N_\mathrm{L} : N_\mathrm{S} = 0.65 : 0.35$ and equal mass.
The interactions between particles are governed by the Weeks-Chandler-Anderson (WCA) potential
\begin{equation}
    U_{ab}(r)=\begin{cases} 4\epsilon_{ab}[(\frac{\sigma_{ab}}{r})^{12}-(\frac{\sigma_{ab}}{r})^6]+\epsilon_{ab}, & r\leq2^{1/6}\sigma_{ab}\\
    0, & r>2^{1/6}\sigma_{ab}
    \end{cases}
    \nonumber
\end{equation}
where $a,b \in {\mathrm{L}, \mathrm{S}}$ and $r$ is the inter-particle distance. The interaction parameters are set as: $\sigma_\mathrm{LL} = 1.0$, $\sigma_\mathrm{SS} = 0.88$, $\sigma_\mathrm{LS} = 0.8$, $\epsilon_\mathrm{LL} = 1.0$, $\epsilon_\mathrm{SS} = 0.5$, and $\epsilon_\mathrm{LS} = 1.5$.

The system is confined within a square box with periodic boundary conditions, with a number density fixed at $N/L^2 = 1.22 \sigma_{LL}^{-2}$. Simulations are carried out in the canonical ($NVT$) ensemble using a Nosé-Hoover thermostat~\cite{nose1984unified}, and the integration time step is set to $0.001$. We begin by equilibrating the system at $k_\mathrm{B}T = 1.0$ for $2 \times 10^4 \tau_0$, where the characteristic collision time is $\tau_0 = 100$ steps. The system is then rapidly quenched to the target temperature, approximately $k_\mathrm{B}T = 0.10$, below which the mean squared displacement (MSD) exhibits a pronounced plateau extending over $10^5 \tau_0$ (Extended Data Fig.~\ref{fig:simualtion}a).

\subsection*{Hierarchical dynamics and relaxations analysis.}

We first calculate the mean square displacement (MSD) as $\langle\Delta r^2(\tau)\rangle/a^2=\frac{1}{N}\sum_{i}\langle[\mathbf{r}_i(t+\tau)-\mathbf{r}_i(t)]^2\rangle_t/a^2$, where $N$ is the number of particles and $a$ is the average interparticle distance. To characterise structural relaxations, we use the self part of the intermediate scattering function: 
\begin{equation}
    F_s(q,\tau)=\langle\frac{1}{N}\sum_{i=1}^Ne^{-i\mathbf{q}\cdot\Delta \mathbf{r}_i(\tau)}\rangle_t\nonumber
\end{equation}
where $\Delta\mathbf{r}_i(\tau)$ is displacement of particle $i$ over a time interval $\tau$ timescale, and $q=\lvert\mathbf{q}\rvert$. In our analysis, we chose $q=2\pi/a$ to probe $\alpha$-relaxation dynamics. The structural relaxation time $\tau_\alpha$ is then defined as the time at which $F_s(q,\tau)$ decays to $e^{-1}$.

The hierarchical dynamics of colloidal systems is directly characterised by the relaxation spectrum, which is obtained by calculating the imaginary part of the dynamic susceptibility $\chi''(\tau^{-1})$, defined as~\cite{scalliet2022thirty,guiselin2022microscopic}:
\begin{equation}
    \chi''(\tau^{-1})=-{\frac{1}{\tau}}\int_0^\infty G(t)\frac{\tau}{1+t^2/\tau^2}\mathrm{d}t\nonumber
\end{equation}
where $G(t)=-\mathrm{d}F_s(q,t)/\mathrm{d}t$ represents the time derivative of the self-intermediate scattering function and reflects the distribution of relaxation times $\tau$.
For better visual comparison, the results shown in Fig.~\ref{fig:Fig2}a are rescaled such that the maxima of $\chi''(\tau^{-1})$ (corresponding to the $\alpha$ peaks) are normalised to the same height.

At the single-particle level, we use the ratio of broken bonds $C_b^i$ to visualise bond-breaking relaxation events in real space. Neighbours of particle $i$ are first identified using Voronoi tessellation analysis. The ratio of broken bonds is then defined as $C_b^i = n_i(t \lvert 0) / n_i(0)$, where $n_i(0)$ is the number of neighbours at the initial time and $n_i(t \lvert 0)$ is the number of those initial neighbours that remain bonded to particle $i$ at time $t$.

\subsection*{Dynamic eigenmodes and spectrum analysis.}

The dynamic eigenmode approach is based on the eigen microstate theory~\cite{hu2019condensation,sun2021eigen}, and serves as a powerful tool for analysing the dynamics and phase transitions of complex systems, including applications in climate systems~\cite{chen2021eigen}, living systems~\cite{li2021discontinuous}, and beyond.

The main procedures are described in the main text. The dimension of the matrix $\mathbf{D}$ is $dN$, where $d$ is the dimensionality of the system ($d = 2$ for our 2D system). In our calculations, the total $t_\mathrm{a}$ period consists of $M \sim 10{,}000$ frames, which exceeds $dN \approx 5{,}000$. We find that both the polarisation vectors $\mathbf{e}_\lambda$ and the eigenvalues $\lambda$ associated with large-$\lambda$ modes ($\nu < \nu_\ast$) are robust with respect to $M$, provided that $M > dN$. Additionally, each displacement interval $\Delta t$ contains at least 200 frames, ensuring that the dynamic eigenmode analysis can effectively resolve different relaxation events occurring at distinct moments in time.

We then diagonalise the symmetric matrix $\mathbf{D}$ to obtain $dN$ eigenvectors $\mathbf{e}_\lambda$ and their corresponding eigenvalues $\lambda$. The diagonal elements $\mathbf{D}_{ii}$ represent the MSD of particle $i$ at the timescale $\Delta t$, i.e., $\langle\Delta\mathbf{r}_i^2(\Delta t)\rangle$. It is important to note that the sum of all eigenvalues equals the trace of the matrix $\mathbf{D}$. Therefore, we have $\sum \lambda = \mathrm{Tr}(\mathbf{D}) = \sum_{i=1}^N \langle \Delta \mathbf{r}_i^2(\Delta t) \rangle = \sum \langle \Delta \mathbf{r}_\lambda^2(\Delta t) \rangle$, where $\lambda = \langle \Delta \mathbf{r}_\lambda^2(\Delta t) \rangle$ quantifies the contribution of eigenmode $\mathbf{e}_\lambda$ to the MSD at the $\Delta t$ timescale.

It should be noted that $\mathbf{D}$ is not mathematically guaranteed to be semi-positive definite. In our calculations, approximately 0.2\% of the eigenmodes exhibit negative $\lambda$. However, these modes have extremely small magnitudes (on the order of $\sim10^{-8}a^2$), and their corresponding eigenvectors $\mathbf{e}_\lambda$ display random spatial patterns with negligible similarity to the actual displacement fields.

The ANOS spectrum of eigenmodes is determined as follows. We introduce a dimensionless parameter $\nu = a^2 / \lambda$ to re-label the eigenmodes, where $a$ is the average interparticle distance and $\lambda$ is the eigenvalue. The accumulated number of states (ANOS) is then calculated as $N(\nu) = \sum_{\nu_\lambda < \nu} \delta(\nu - \nu_\lambda)$, where $\delta(x)$ is the Dirac delta function. The reduced density of states (DOS) spectrum, shown in Extended Data Fig.~\ref{fig:basic}a, is derived based on the Debye scaling relation. Assuming an effective ``frequency'' $\omega\propto\sqrt{\nu}$, we obtain the standard Debye scaling $D(\omega) \sim \omega^{d-1} \sim \omega$ for a 2D system ($d = 2$). Note that this assumed relationship between spatial amplitude (related to $\nu$) and vibrational frequency $\omega$ may hold in low-temperature solid-like systems. 

Based on this relation, the reduced DOS spectrum can be expressed as:
\begin{equation}
    D(\sqrt{\nu})/\sqrt{\nu}=\frac{2}{dN}\frac{\mathrm{d}N(\nu)}{\mathrm{d}\nu}. \nonumber
\end{equation}
From this, it follows that if the DOS scales as $D(\omega) \sim \omega^{k'}$, then the ANOS follows $N(\nu) \sim \nu^k$, where $k = (k'+1)/2$. Accordingly, a DOS scaling with $k' = 4,\ d,\ d-2$ (for $d = 2$) corresponds to ANOS segments with $k = 5/2,\ 3/2,\ 1/2$.

\subsection*{Covariance matrix analysis.}

The covariance matrix is a standard method for extracting vibrational modes and spectra in experimental systems~\cite{chen2010low}. It is defined as $C_{ij} = \langle [\mathbf{r}_i(t) - \bar{\mathbf{r}}_i] \cdot [\mathbf{r}_j(t) - \bar{\mathbf{r}}_j] \rangle_t$, where $\bar{\mathbf{r}}_i$ denotes the equilibrium position of particle $i$, and the matrix captures fluctuations around these positions. 
When particle motions can be approximated as harmonic—such as at $T \to 0$ or over short timescales without structural rearrangement—the covariance matrix can be related to the dynamical Hessian matrix via $\mathbf{H} = \frac{k_B T}{m} \cdot \mathbf{C}^{-1}$. The eigenvectors and eigenvalues of $\mathbf{H}$ correspond to the vibrational modes and their associated frequencies.
In this work, we compare the conventional covariance matrix $\mathbf{C}$ with the dynamic displacement correlation matrix $\mathbf{D}$ introduced in our approach, using a simulation system as shown in Extended Data Fig.~\ref{fig:simualtion}.

\subsection*{Spatial distribution and effective ``frequency'' of eigenmodes analysis.}

The spatial distribution of $\mathbf{e}_\nu$ is characterised by the participation ratio $e$:
\begin{equation}
    e=\frac{(\sum_{i=1}^N\lvert\mathbf{e}_\nu^i\rvert^2)^2}{N\sum_{i=1}^N(\mathbf{e}_\nu^i\cdot\mathbf{e}_\nu^i)^2} \nonumber
\end{equation}
where $\mathbf{e}_\nu^i$ is the polarisation vector of particle $i$ in the eigenmode $\mathbf{e}_\nu$. The participation ratio quantifies the degree of spatial localisation: smaller values of $e$ indicate stronger localisation. In particular, $e \to 1/N$ corresponds to highly localised modes, while $e = 1$ indicates uniform translational modes. A value of $e = 2/3$ is typical for phonon-like modes.
In our glassy system, quasi-elastic modes exhibit phonon-like spatial distributions with $e \approx 2/3$. Random-like noise modes, whose vector amplitudes follow a Gaussian distribution, show $e \approx 0.5$. In contrast, string modes are more spatially heterogeneous, especially near $\phi_\mathrm{g}$, as indicated by small values $e < 0.4$ at $\phi = 0.83$ (Extended Data Fig.~\ref{fig:basic}b).

For each eigenmode, we estimate its effective ``frequency'' $\omega$ by Fourier transforming its activation curve $c_\nu(t,\delta t)$. The power spectrum of Fourier transformation extracts the main frequencies of eigenmodes. For the spatial information, we project the dynamic eigenmode $\mathbf{e}_\nu$ onto vibrational mode(VM) $\mathbf{e}_\omega$ obtained through Hessian matrix, and evaluate their similarity by $C_{\nu,\omega}^2=(\mathbf{e}_\nu\cdot\mathbf{e}_\omega)^2$. We also sum up the largest six $C_{\nu,\omega}^2$ of a single $\mathbf{e}_\nu$ to evaluate the extend of mode mixing (Extended Data Fig.~\ref{fig:perturbation}).

At finite temperature, the mode mixing caused by thermal fluctuations and anharmonic effect leads to the multi-frequency characteristic of dynamic eigenmodes. Under weak anharmonicity at $T\to 0$, a single eigenmode is composed of few VM, and the frequencies of eigenmodes preserve a well-defined $\omega\sim\nu^{1/2}$ relation (Extended Data Fig.~\ref{fig:perturbation}a). However, around $T_\mathrm{g}$, a single eigenmode represents the co-activation of multiple VM with broader frequency distribution. Its effective frequencies $\omega$ therefore become much lower than the VM (Extended Data Fig.~\ref{fig:perturbation}b). However, we note that the quasi-elastic modes [(i)] are composed of less VM, indicating that they may be more ``elastic'' than the other types of modes.

\subsection*{Softness parameter calculation.}

Soft spots in our systems are identified using the softness parameter $S_i$, defined as $S_i = \sum_{\nu < \nu_\ast} \lvert \mathbf{e}_\nu^i \rvert^2$, where $\mathbf{e}_\nu^i$ is the polarisation vector of particle $i$ in mode $\nu$. Specifically, the results shown in Fig.~\ref{fig:Fig3}c use a more focused definition: $S_i = \sum_{\nu_0 < \nu < \nu_t} \lvert \mathbf{e}_\nu^i \rvert^2$, which isolates the contribution from intermediate modes in region (ii).
This analysis is inspired by the local Debye-Waller factor defined in terms of vibrational modes~\cite{widmer2008irreversible,widmer2009localized}, which connects the system's linear response to its vibrational properties. In our study, we extend this approach to quantify the probability of relaxation through string-like motions, as captured by the string modes in categories (ii), (iii), and (iv).

\subsection*{Dynamic heterogeneity analysis.}

We use four-point susceptibility $\chi_4(t)$ to characterise the dynamic heterogeneity in the system. The calculation begins with computing the structural self-overlap function as $Q_s(a_p,t) = \frac{1}{N} \sum_{i=1}^N \exp(-\Delta r_i^2(t) / 2a_p^2)$~\cite{dauchot2005dynamical,gokhale2014growing,berthier2011dynamical}, where $\Delta r_i^2(t)$ is the squared displacement of particle $i$ over time $t$, and $a_p$ is a probing length scale.
The four-point susceptibility is then obtained from the variance of $Q_s(a_p, t)$:
\begin{equation}
    \chi_4(t)=N(\langle Q_s^2(a_p,t)\rangle-\langle Q_s(a_p,t)\rangle^2) \nonumber
\end{equation}
We set the probing length to $a_p = 0.3a$, which corresponds to the typical amplitude of particle relaxations in our system and maximises the peak of $\chi_4(t)$~\cite{gokhale2014growing}. The peak timescale of $\chi_4(t)$ coincides with the structural relaxation time $\tau_\alpha$, indicating that the dynamics are most heterogeneous around this timescale~\cite{berthier2011dynamical}.

\bibliographystyle{naturemag_noURL}
\bibliography{main}

\vspace{0.8cm}
\noindent
{\bf Data availability}

\noindent
The data that support the findings of this study are available from the corresponding author upon reasonable request.

\noindent
{\bf Code Availability}

\noindent
The codes used in this study are available from the corresponding authors upon reasonable request.

\vspace{1cm}
\noindent
{\bf Acknowledgments} We acknowledge the National Natural Science Foundation of China (Nos. 12425503, 12174071, 12035004 and 12135003), the Innovation Program of Shanghai Municipal Education Commission (No. 2023ZKZD06), and Shanghai Pilot Program for Basic Research-FuDan University (No. 22TQ003). H.T. acknowledges a Grant-in-Aid for Specially Promoted Research (JP20H05619) from the Japan Society of the Promotion of Science (JSPS). 

\noindent
{\bf Author contributions}
H. T. and P. T. conceived and supervised the research. W. S. and Y. C. performed the experiments. Y. Z. performed the simulations. All authors made the data analysis. W. S., W. J., H. T. and P. T. wrote the paper.

\noindent
{\bf Competing interests} The authors declare no competing interests.

\noindent
{\bf Additional Information}
\noindent

\clearpage
\newpage

\clearpage
\setcounter{figure}{0}
\renewcommand{\figurename}{\textbf{Extended Data Fig.}}

\begin{figure*}[ht]
\includegraphics[width=0.93\linewidth]{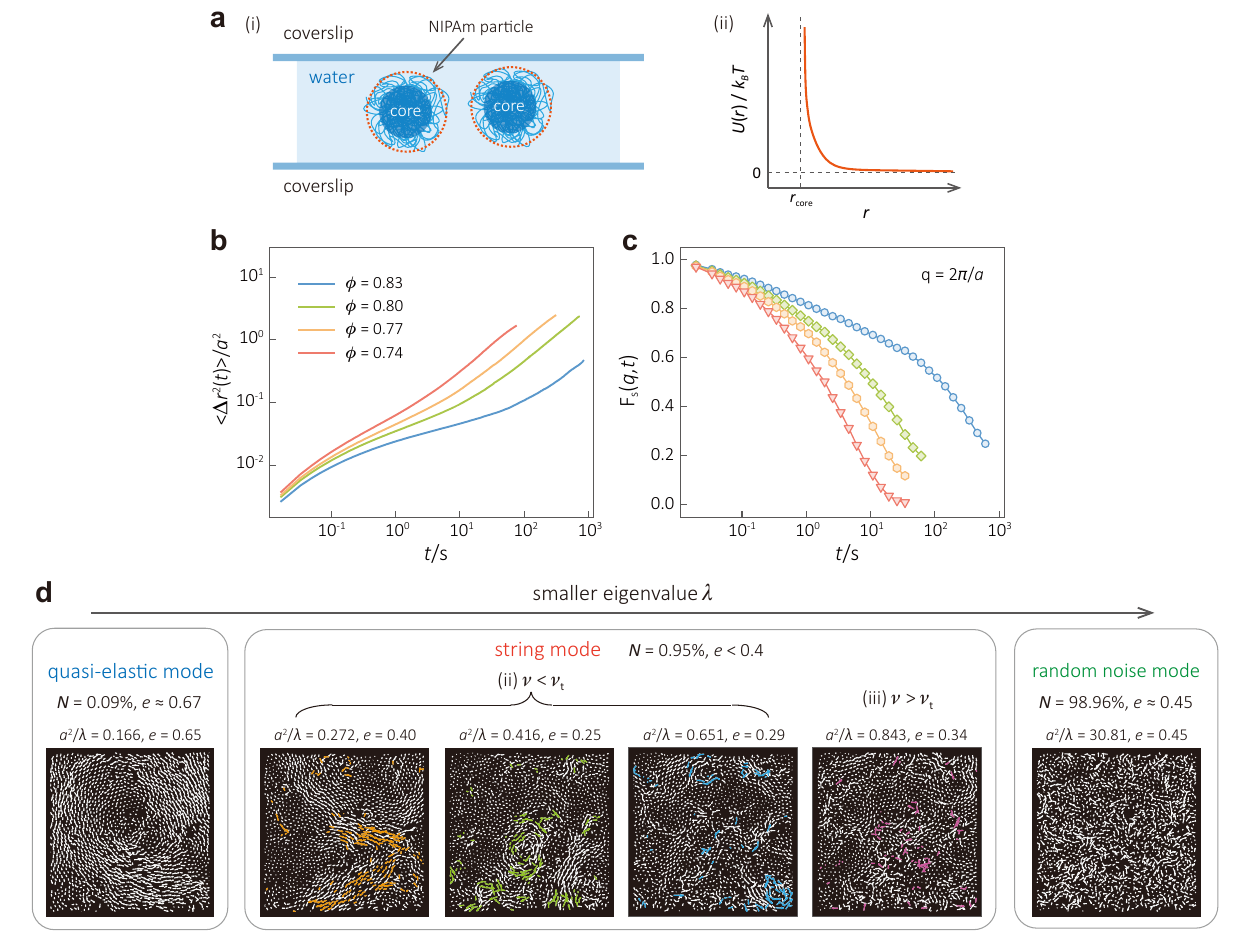}
\caption{\textbf{Properties and dynamic eigenmodes of NIPAm colloidal system.} 
$\mathbf{a}$, Schematics of NIPAm colloidal system (i) and the pair interaction between particles (ii). NIPAm colloids exhibit a typical core-corona structure and interact with each other through short-range hard-core repulsion~\cite{chen2011measurement}.
$\mathbf{b}$, Mean square displacement (MSD). The dynamics of the particles consist of short-time inside-cage rattlings (evidenced by the plateau) and long-time sub-diffusive behaviour. 
$\mathbf{c}$, The intermediate scattering function $F_s(q,t)$ with $q=2\pi/a$. The characteristic structural relaxation time $\tau_\alpha$ is estimated by $t$ where $F_s(q,t)=e^{-1}$.
$\mathbf{d}$, Polarisation vectors of several dynamic eigenmodes of $\phi=0.83$ system with $\Delta t = 0.1\tau_\alpha$ and $t_{\rm a} = 2.5\tau_\alpha$. The total number of eigenmodes is $dN=4,958$. The corresponding ANOS spectrum is shown in Fig.~\ref{fig:Fig1}c. Left, the quasi-elastic modes (i) with extended vortex-like structure reflect the emergent quasi-elasticity as the system approaches $\phi_\mathrm{g}$. In the middle, the string modes composed of string-like elements. These modes capture the string-like collective intermediate relaxations and have the highest contribution to long-time $\alpha$-relaxations. The coloured strings contribute significantly to the real dynamics (Fig.~\ref{fig:Fig1}h), as shown in Fig.~\ref{fig:Fig1}i. Right, the random noise modes that represent the isolated inside-cage random motions under thermal fluctuations.
\label{fig:NIPAm}
}
\end{figure*}

\newpage
\begin{figure*}[ht]
\includegraphics[width=\linewidth]{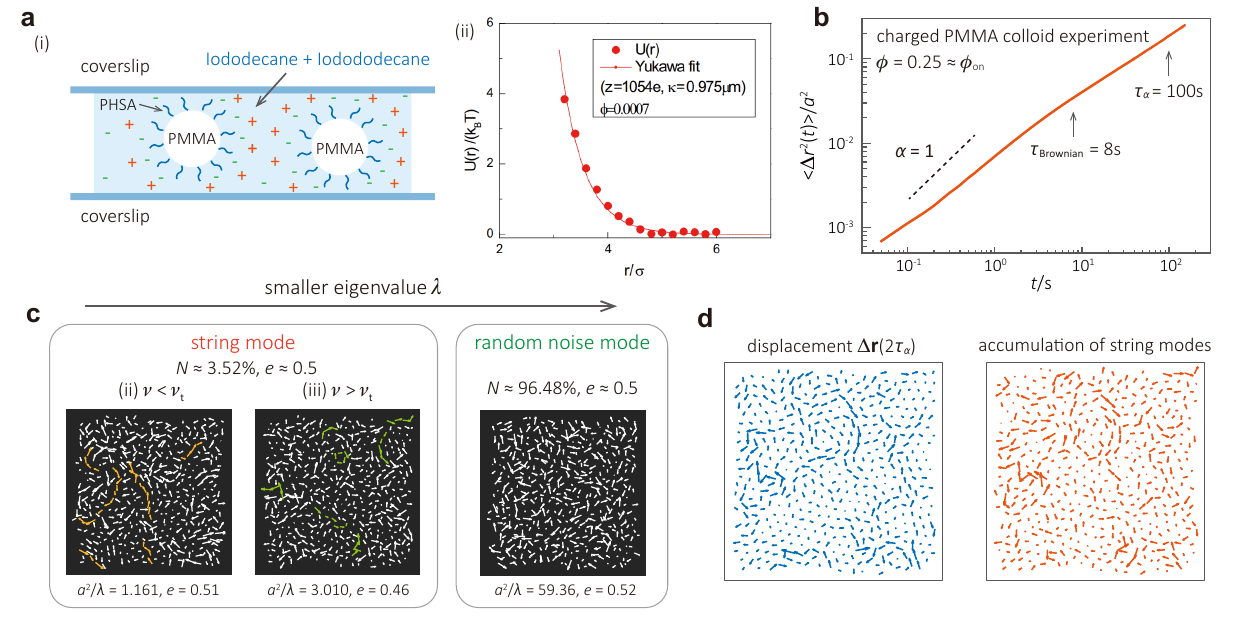}
\caption{\textbf{Properties and dynamic eigenmodes of charged PMMA colloidal system ($\phi=0.25\approx\phi_\mathrm{on}$).}
$\mathbf{a}$, Schematic of the charged PMMA colloidal system (i) and the pair interaction between particles (ii). The PMMA particles are coated with PHSA and are negatively charged in the solution. The electric double layer leads to long-range hard-core repulsive interactions consistent with the Yukawa fit (ii). $\mathbf{a}$(ii) is adapted from the reference~\cite{tan2012understanding}.
$\mathbf{b}$, Mean square displacement (MSD).
$\mathbf{c}$, Dynamic eigenmodes with $\Delta t = 0.1\tau_\alpha$ and $t_{\rm a} = 2.0\tau_\alpha$. The total number of eigenmodes is $1,000$. There is no quasi-elastic modes (i) in this system. The string modes become extended ($e\approx0.5$) but still consist of string-like elements.
$\mathbf{d}$, Comparison between real displacement during $2.0\tau_\alpha$ (left) and the projected displacement from modes (ii) and (iii) (right). These results demonstrate the effectiveness of DEA across glassy systems with different $\phi$ and interaction potentials.
\label{fig:PMMA}
}
\end{figure*}

\newpage
\begin{figure*}[ht]
\includegraphics[width=0.95\linewidth]{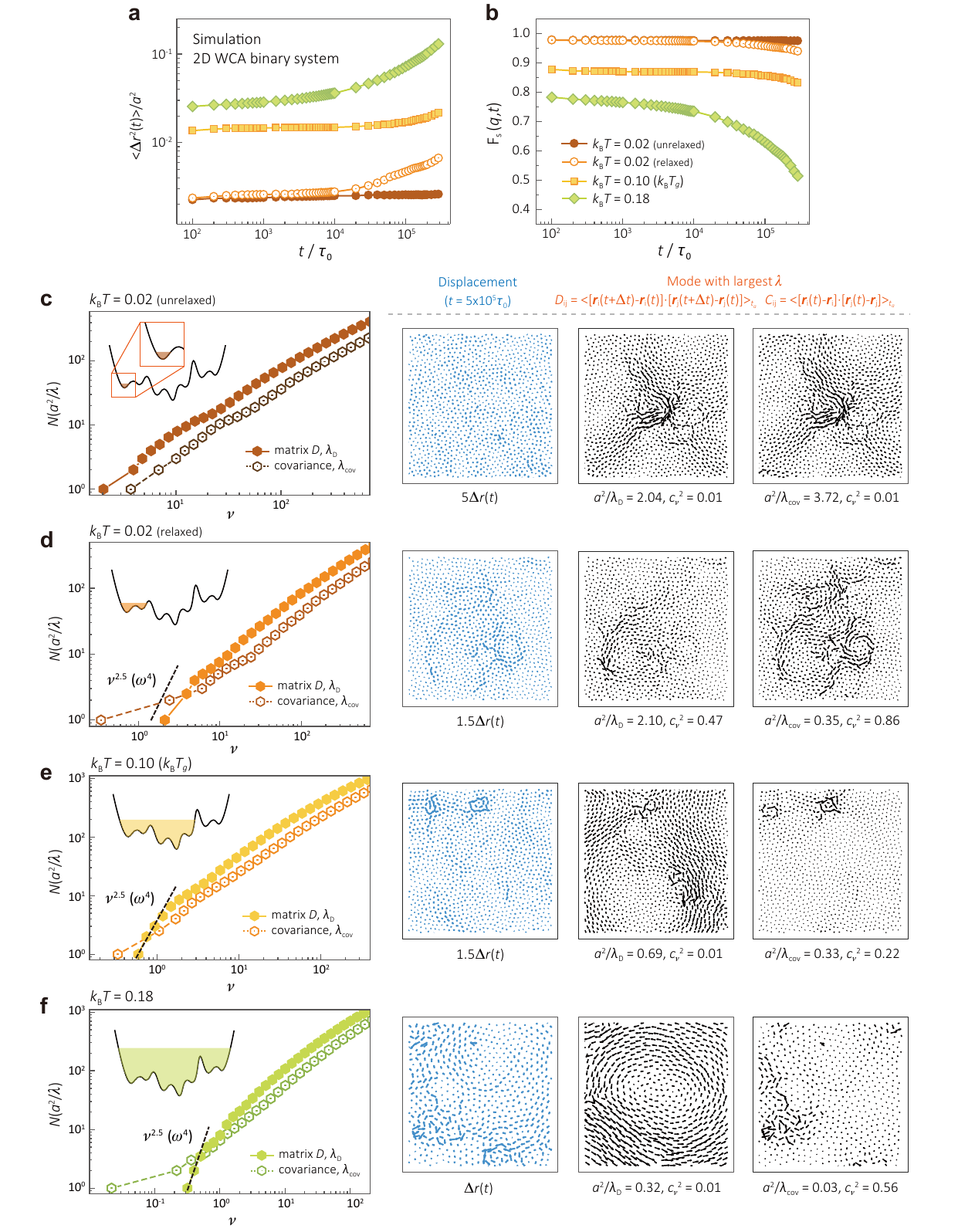}
\caption{
\label{fig:simualtion}
}
\end{figure*}

\clearpage
\newpage
\noindent
\textbf{Extended Data Fig. 3: Comparison of DEA and covariance matrix method through MD simulations.} 
$\mathbf{a}$-$\mathbf{b}$, MSD and $F_s(q,t)$ of glassy systems below, at, and above $T_{\rm g}$. $\tau_0$ is the interval of particles' collisions. The glass transition temperature of this system is $k_\mathrm{B}T_{\rm g} = 0.10$, where the plateau of MSD extends to $10^5 \tau_0$.
$\mathbf{c}$-$\mathbf{f}$, ANOS spectra and smallest-$a^2 / \lambda$ modes of matrix $\mathbf{D}$ and covariance matrix $\mathbf{C}$. We use $\Delta t=2\times10^4\tau_0$ and $t_\mathrm{a}=5\times10^5\tau_0$ for calculating $\mathbf{D}$, and same $t_\mathrm{a}$ for calculating $\mathbf{C}$ (this $t_\mathrm{a}$ contains $10^4$ frames). 
In the low-temperature system without structural rearrangement, these two matrices both reveal the vibrational motions with similar spectra and modes ($\mathbf{c}$). Eigenmodes with the largest $\lambda$ are typical quasi-localised vibrational modes consisting of localised soft spots and vortex-like background. For systems with structural relaxations ($\mathbf{d}$-$\mathbf{f}$), the covariance matrix fails to distinguish different dynamic behaviours, as the largest-$\lambda$ eigenmode exhibits high similarity with the displacements. In this region, matrix $\mathbf{D}$ is capable for separating different motions. Around $T_\mathrm{g}$ ($\mathbf{e}$-$\mathbf{f}$), we observe the quasi-elastic modes which correspond to the $\nu^{2.5}$ scaling region on ANOS spectrum. These quasi-elastic modes may stem from the excitation of stable passive mode hybridised with extended modes~\cite{ji2019theory}.


\newpage
\begin{figure*}[ht]
\includegraphics[width=0.8\linewidth]{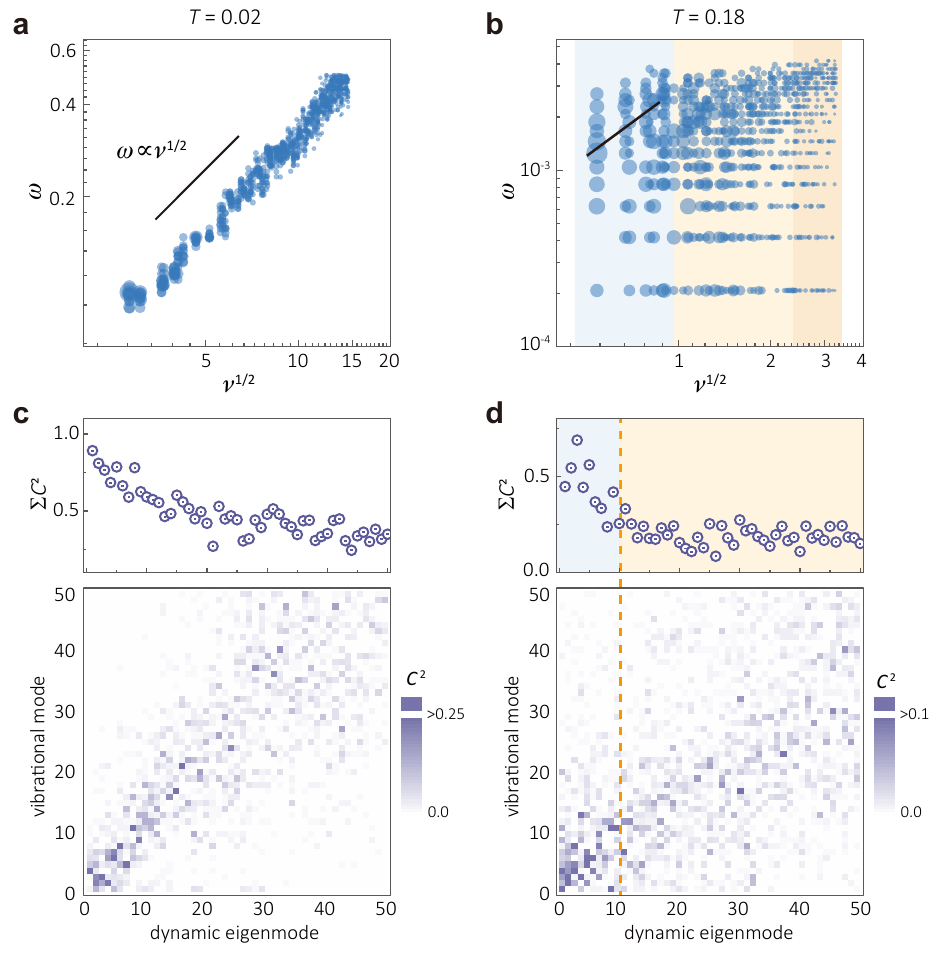}
\caption{\textbf{Comparing the dynamic eigenmodes and vibrational modes obtained from Hessian matrix.}
\textbf{a}-\textbf{b}, Relation between $\nu$ and the effective ``frequency'' $\omega$ of dynamic eigenmodes. For each eigenmode, we show the top 10 frequencies on its Fourier power spectrum. At $T=0.02$, the frequencies of eigenmodes approximately follow $\omega\sim\nu^{1/2}$. At $T=/0.18$, the mode mixing caused by nonlinear effect leads to a multi-frequency characteristics and the much lower frequencies $\omega$. The corresponding ANOS spectra are shown in Extended Data Fig. \ref{fig:simualtion}\textbf{c} and \textbf{f}.
\textbf{c}-\textbf{d}, Similarity between small-$\nu$ dynamic eigenmodes and low-frequency vibrational modes (VM) of Hessian matrix. The upper figures show the summation $\Sigma C^2$ over $C^2$ of six VM with largest $C^2$. The eigenmodes at $T=0.02$ and quasi-elastic modes [(i)] at $T=0.18$ are similar to the VM, supporting the perturbation framework.
\label{fig:perturbation}
}
\end{figure*}

\newpage
\begin{figure*}[ht]
\includegraphics[width=\linewidth]{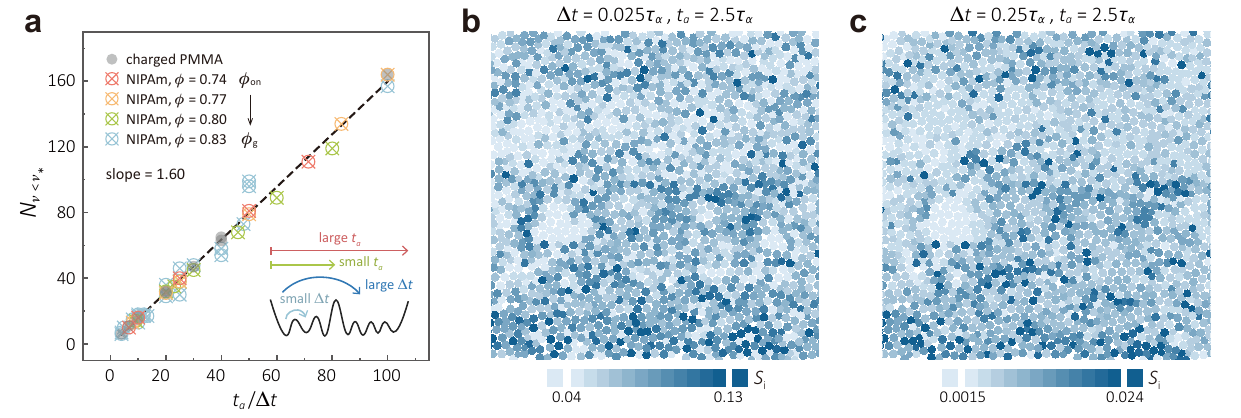}
\caption{\textbf{Influence of matrix parameters $\Delta t$ and $t_{\rm a}$.} 
$\mathbf{a}$, Relations between the total number of modes in (i)-(iv) ($\nu<\nu_*$) and the ratio of parameters $t_{\rm a} / \Delta t$. Based on the definition of matrix $\mathbf{D}$, $\Delta t$ controls the ``resolution'' of detection on the potential energy landscape (PEL), while $t_{\rm a}$ decides the total range (see the inset schematic). Since the PEL contains multiple inherent basins, $N_{\nu<\nu_*}$ is proportional to both $t_{\rm a}$ and $1 / \Delta t$.
$\mathbf{b}$-$\mathbf{c}$, Softness parameters $S_i$ calculated with different $\Delta t$. With smaller $\Delta t$, we capture more details of intermediate relaxations within a smaller typical timescale. Additionally, the similarity among $\mathbf{b}$, $\mathbf{c}$, and Fig.~\ref{fig:Fig1}e shows the effectiveness and robustness of DEA.
\label{fig:Si}
}
\end{figure*}

\newpage
\begin{figure*}[ht]
\includegraphics[width=\linewidth]{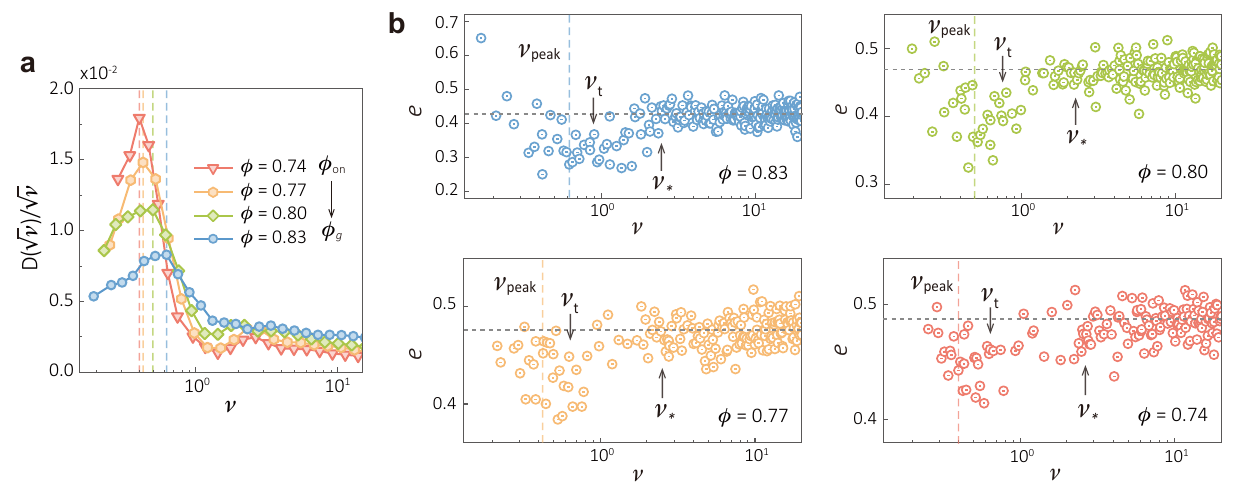}
\caption{\textbf{Reduced density of states (DOS)  and participation ratio of the modes in four NIPAm systems towards glass transition.} 
$\mathbf{a}$, Reduced density of states (DOS) spectra of dynamic eigenmodes (Methods). Modes close to $\nu_t$ appear as a peak denoted by vertical lines. As $\phi$ increases, the peak shifts upwards and decreases in height.
$\mathbf{b}$, Participation ratio $e$ of dynamic eigenmodes in four systems. As $\phi$ decreases, the quasi-elastic modes disappear when $\phi\leq0.80$ and the string modes become more extended. Around $\nu_\mathrm{peak}$, the string modes (modes in (ii) since $\nu_\mathrm{peak}<\nu_t$) are little quasi-localised, with $e$ forming a dip. Additionally, string modes in (iii)-(iv) ($\nu > \nu_t$) are more extended than modes in (ii) ($\nu < \nu_t$).
\label{fig:basic}
}
\end{figure*}

\newpage
\begin{figure*}[ht]
\includegraphics[width=\linewidth]{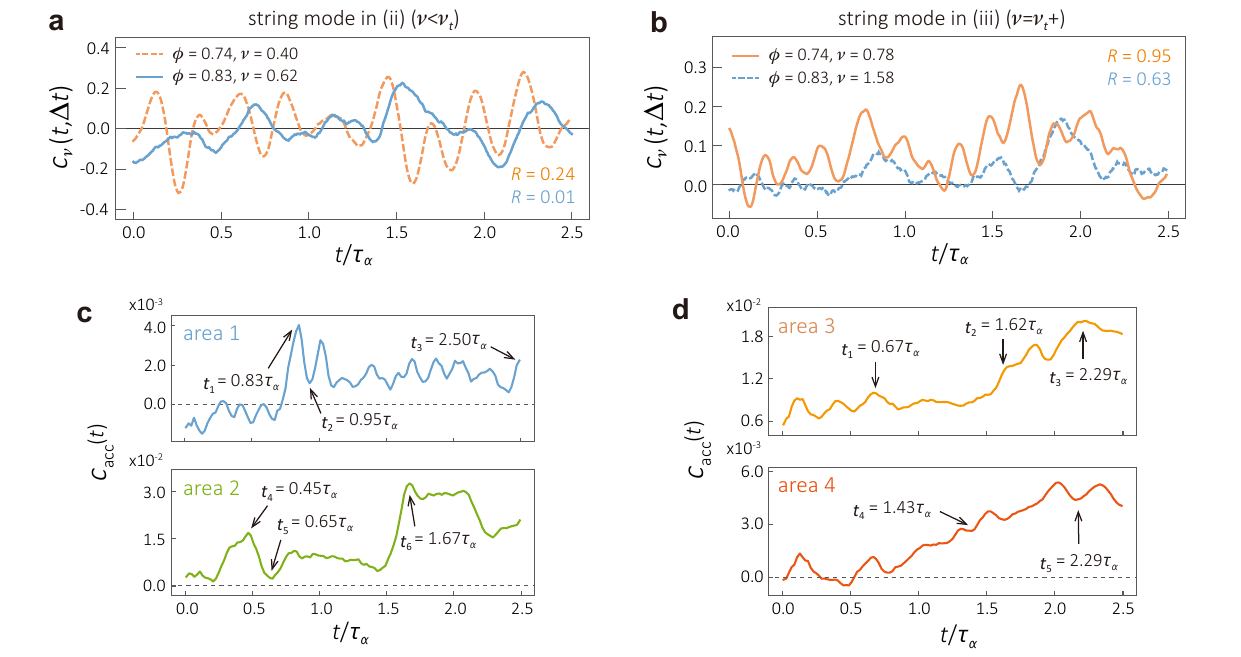}
\caption{
\textbf{Reversibility of string modes in (ii) and (iii).} 
$\mathbf{a}$-$\mathbf{b}$, Typical activation curves of the reversible modes in (ii) ($\mathbf{a}$, $\nu = \nu_\mathrm{peak} < \nu_t$) and irreversible modes in (iii) ($\mathbf{b}$, $\nu > \nu_t$). The curves are obtained by calculating $c_\nu(t, \Delta t)=\mathbf{e}_\nu\cdot\Delta\mathbf{r} / \lvert \Delta \mathbf{r} \rvert$ over time $t$. Curves of the reversible modes (ii) oscillate around $0$ (marked by gray solid lines) as shown in $\mathbf{a}$. In contrast, the curves of irreversible modes (iii) is almost entirely above $0$ as shown in $\mathbf{b}$. The solid blue curve in $\mathbf{a}$ corresponds to the mode (ii) in Fig.~\ref{fig:Fig2}d, while the solid orange curve refers to the mode (iii) in Fig.~\ref{fig:Fig2}e.
$\mathbf{c}$-$\mathbf{d}$, Comparisons of distinct string-like excitations in modes (ii) and modes (iii): the large-amplitude reversible hopping motions ($\mathbf{c}$, refer to strings in Fig.~\ref{fig:Fig2}d), and the unidirectional flows ($\mathbf{d}$, strings in Fig.~\ref{fig:Fig2}e). Here $c_{\mathrm{acc}}(t)=\sum_i\mathbf{e}^i_\nu \cdot \Delta\mathbf{r}_i / \lvert \Delta \mathbf{r}_i \rvert$ with $i$ being particles in the string and $\Delta\mathbf{r}_i=\mathbf{r}_i(t)-\mathbf{r}_i(0)$. The string-like motions in the same eigenmode occur at similar moments. The plateau between the rapid increases or drops of $c_{\mathrm{acc}}(t)$ indicates that excitation of strings are separated by the inside-cage vibrations.
\label{fig:reversibility}
}
\end{figure*}

\newpage
\begin{figure*}[ht]
\includegraphics[width=\linewidth]{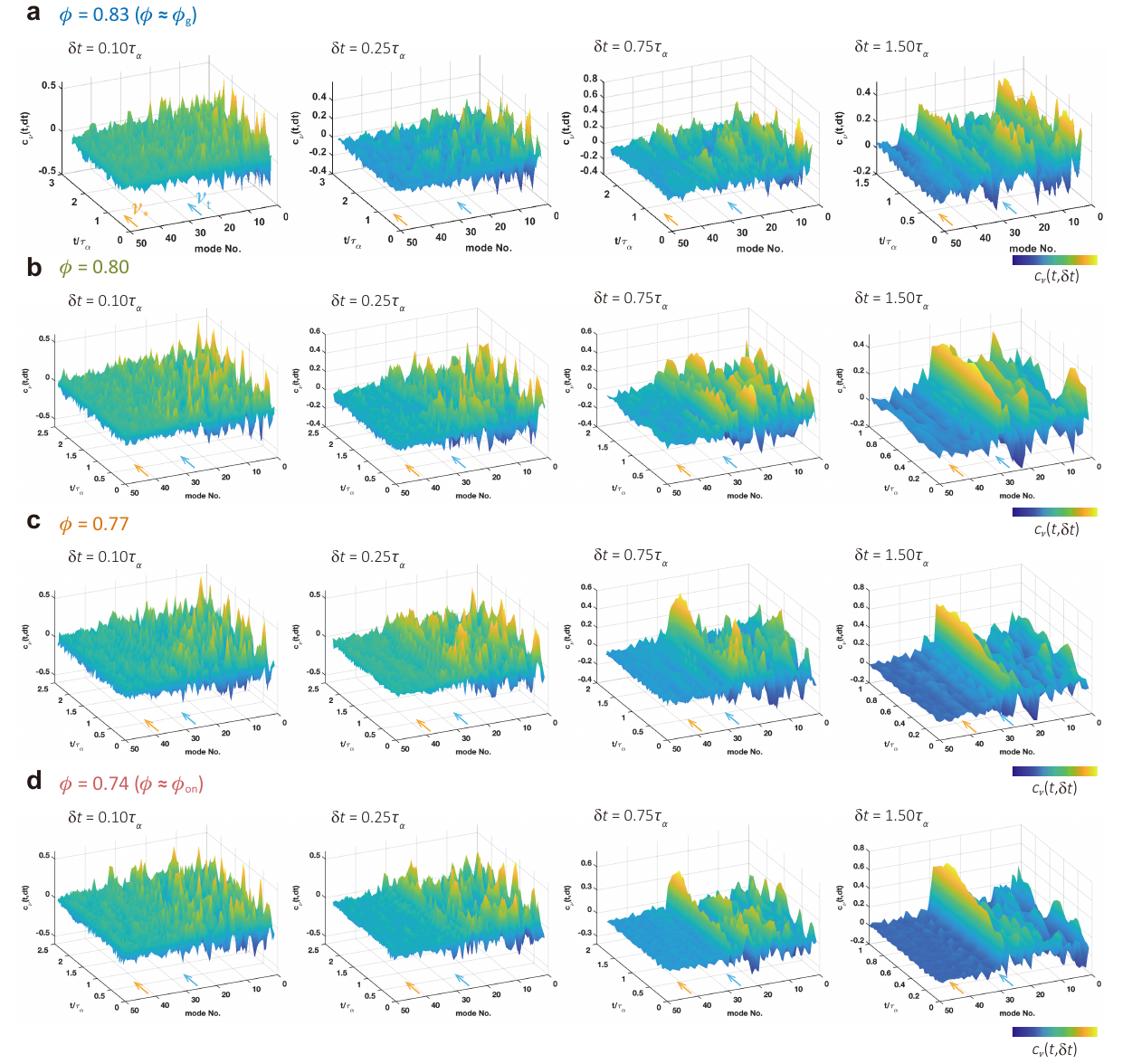}
\caption{\textbf{Dynamics at different timescales projected to string modes (ii)-(iv).}  $\phi = 0.83 \approx \phi_g$ in $\mathbf{a}$, $\phi = 0.80$ in $\mathbf{b}$, $\phi = 0.77$ in $\mathbf{c}$, and $\phi = 0.74 \approx \phi_{on}$ in $\mathbf{d}$. We plot the parameter $c_\nu(t, \delta t)$ in the form of  $(N_\nu, t/\tau_{\alpha}, c_\nu)$ $3d$ landscapes. We allocate the Mode No. $N_\nu$ as the $N_\textit{th}$ mode with an increase of $\nu$. The characteristic values, $\nu_t$ and $\nu_*$, are marked with blue and orange arrows, respectively. At larger $\delta t$, $c_\nu$ of modes (ii) and (iii) increases, transforming the landscape morphology from a more uniform type (left-first panels) to ones with several pronounced peaks (right-second and right-first panels). Dynamics projected to modes (ii) exhibit bell-like peaks along the time axis (see those in $\delta t=0.75$ and $1.0\tau_\alpha$). In contrast, dynamics projected to modes (iii) appear as plateaus with growing height at longer times. Figure~\ref{fig:Fig3}a provides the side view profiles of these landscapes from the ``Mode No.''axis.
\label{fig:landscape}
}
\end{figure*}

\end{document}